\documentclass{aa}
\usepackage{graphicx}
\usepackage{txfonts}
\usepackage{lscape} % PIERRE: POUR AVOIR UNE TABLE EN PAYSAGE (OBSERVATIONS)

\begin{document}

%\documentstyle[11pt,aasms4]{article}

% Plots from globularcluster2.pro, ....
% Data from data.tex, ...
\title{Surface convection and red giants radii measurements} 

\author{L. Piau\inst{1} \and P. Kervella\inst{2} \and S. Dib\inst{3} \and P. Hauschildt\inst{4}} 
\offprints{laurent.piau@cea.fr}
\mail{Pierre.Kervella@obspm.fr}
\institute{
Service d'astrophysique, CEA Saclay, 91191, Gif-sur-Yvette France
\and
LESIA, Observatoire de Paris, CNRS\,UMR\,8109, UPMC, Universit\'e Paris Diderot, 5 place
Jules Janssen, 92195 Meudon, France
\and
Astrophysics Group, Imperial College of Science, Technology and Medicine, London SW7 2AZ, United Kingdom
\and
Hamburger Sternwarte, Gojenbergsweg 112, 21029 Hamburg, Germany
}
\date{Received ; Accepted}

\abstract{ The phenomenological models of convection 
use characteristic length scales they do not determine but 
that are chosen to fit solar or stellar observations. We investigate if changes of these
length scales are required between the Sun and low mass stars on the red giant branch (RGB). 
The question is addressed jointly in the frameworks of the mixing length theory and of 
the full spectrum of turbulence model. For both models, the convective length 
scale is assumed to be a fixed fraction of the local pressure scale height.

We use constraints coming from the observed effective temperatures 
and linear radii {\it independently}. We rely on a sample of 38 
nearby giants and subgiants for which surface 
temperatures and luminosities are known accurately and the radii are determined through 
interferometry to better than 10\%. For the few cases 
where the stellar masses were determined by asteroseismological measurements, we 
computed dedicated models. First we calibrate the solar models. Then, with the same
physics, we compute RGB models for masses between $\rm 0.9 M_{\odot}$
and $\rm 2.5 M_{\odot}$ and metallicities ranging from
$\rm [Fe/H]=-0.34$ to solar. The evolution is followed up to $\rm 10^3\,L_{\odot}$.
A special attention is given to the opacities and to the non grey atmosphere models 
used as boundary conditions for which the model of convection is the same 
as in the interior. 

We find that for both the mixing length theory and the 
full spectrum of turbulence model the characteristic solar length 
scale for convection has to be slightly reduced to fit the lower
edge of the observed RGB. The corresponding models also better 
match the expected mass distribution on the
RGB and are in better agreement to the seismic constraints. 
These results are robust whether effective temperatures determined spectroscopically 
or radii determined interferometrically are used.

\keywords{Physical data and processes : convection.
    Stars individual : low mass -- evolution -- red giants.
    Techniques: high angular resolution
    }
}

\maketitle

%\keywords{Physical data and processes : convection.
%    Stars individual : low mass -- evolution -- red giants.
%    Techniques: high angular resolution
%    }

\section{Introduction}\label{sec1}

The modelling of solar and stellar surface convection requires hydrodynamical radiative 
transfer computations (Nordlund et al.~\cite{aake09}). These tridimensional calculations are extremely time consuming
because convection zones motions are characterized by short time scales 
and enormous turbulence.
%In the current
%state of the art the modelling does not go down to the dissipative scales of the 
%flow but uses subgrid models instead (). Moreover the radiative transfer around the surface of a star
%can neither be addressed within the optically thin nor the optical thick approximations. 
%This obliges us to simplify the problem by using methods that save 
%computational power and time such as the multi group method (). 
Therefore, in spite of the current computational power, the direct modelling of convection does not rely 
on first principles only but still necessitates simplifying assumptions. 

Direct observations can constrain the properties 
of outer convection in stars.
The Sun's radius is routinely used to calibrate the
main free parameter of the mixing length theory, $\rm \alpha_{mlt}$,
the characteristic length scale of convection in this model being generaly
$\rm \alpha_{mlt} H_p$ where $\rm H_p$ denotes the local pressure scale heigth. 
It is customary to consider that the solar $\rm \alpha_{mlt}$
applies all the way through stellar evolution and
some recent studies have suggested this to be a valid hypothesis 
for low-mass red giant stars (Ferraro et al.~\cite{ferraro06}; 
VandenBerg et al.~\cite{vdb08}). 
However hydrodynamical convection calculations (Ludwig et al.~\cite{ludwig99}) 
show that slightly different $\rm \alpha_{mlt}$ should be used depending on the 
effective temperature (hereafter $\rm T_{eff}$) and the surface gravity (hereafter log g).
Furthermore, the solar $\rm \alpha_{mlt}$ appears inappropriate to 
describe outer convection in red giants in the $\rm 3-20 \, M_{\odot}$ 
mass range (Stothers \& Chin~\cite{stothers97}) 
and during the final stages of stellar evolution (D'Antona \& Mazzitelli~\cite{dantona96}).
This is one reason why it is interesting to check what the
nearby Galactic red giants can tell us about the mixing length theory (hereafter MLT). 
Besides the MLT there is another local treatment of convection
that is increasingly used: the full spectrum of 
turbulence model (hereafter FST, Canuto, Goldman \& Mazzitelli~\cite{cgm96}).
This approach is more physically consistent than the MLT and shows better 
agreement to numerous sets of observations (Mazzitelli~\cite{mazzitelli99} and 
references therein, Samadi et al.~\cite{sam06}).

As red giants and solar surface conditions ($\rm T_{eff}$, log g) 
strongly differ, they offer an adequate
opportunity to check the MLT and the FST empirically.
Such analyses were made in the recent years.
However in the case of the nearby red giants,
the usual observational constraints, absolute luminosity
and $\rm T_{eff}$ are now supplemented by 
direct interferometric radii measurements
and the new asterosesimic constraints.
In this work we aim at using these new data to estimate $\rm \alpha_{mlt}$
and $\rm \alpha_{cgm}$ and their possible variation from 
the Sun to Galactic red giants. The analysis is 
made for both convection models using constraints 
on effective temperatures and radii independently.
In section \ref{sec2} we give the physical inputs to our stellar evolution
code. We describe in detail the treatment of 
convection and the atmosphere modelling.
In section \ref{sec3} we present and discuss the observation sample.
The calibration of the convective 
length scales relying on the Sun is made in section 
\ref{sec4}. Section \ref{sec5} examines the 
MLT and the FST model
of convection under the constraints coming 
from the red giant branch stars.
We discuss our results and conclude in section \ref{sec6}.

\section{The evolution code and the models}\label{sec2}

We use the CESAM code (Morel~\cite{morel97}), an hydrostatic one dimensional 
stellar evolution code\footnote{A list of the scientific publications using CESAM is available 
at http://www.oca.eu/morel/articles.html}. However the standard version of
CESAM is significantly modified here: to allow the computation
of the stellar structure in the late stages of evolution we changed
the usual integration variables of CESAM (see appendix 1).
The evolution is computed from the zero age main sequence
up to a luminosity of $\rm 10^3 L_{\odot}$ on the red giant branch (hereafter RGB)
for stars less massive than $\rm 2 M_{\odot}$. For 
more massive stars, the models are evolved to the end of
helium core burning on the early asymptotic giant branch (hereafter eAGB).
The nuclear reaction rates are based on the 
NACRE compilation (Angulo et al.~\cite{angulo99}).
The network includes the proton-proton chains, the 
CNO cycle, and 3\rm $\alpha$ and $\rm ^{12}C(\alpha,\gamma)^{16}O$ reactions. 
The neutrino losses are computed according to the analytical 
fits of Itoh et al.~(\cite{itoh96}).
Microscopic diffusion is taken into account in the solar
calibration models (\S \ref{sec4}) and more generally for all the models
down to an effective temperature of 5000 K.
This temperature roughly corresponds to the maximum extent 
in mass of the outer convection 
zone near the base of the RGB, the so-called first dredge up. 
Then because of the deep convective mixing, the subsequent evolution
of diffusive and non diffusive models become 
undistinguishable\footnote{However we recall that even if the diffusive and non diffusive 
models reach the same RGB they do not do so at the same age because
the main sequence downward diffusion of helium and heavy elements
speeds up the evolution on that stage.}  
(see Salaris et al.~(\cite{salaris02}) and references herein) 
and diffusion is not taken into account below $\rm T_{eff}=5000 K$.
 
The calculation of solar radius or RGB/eAGB radii models critically depends on the physical 
assumptions and inputs used in the outer layers. There are four of them. 

{\it i) The opacities}:
The low temperature opacities dominated by metallicity 
effects are critical in determining
the effective temperature 
of giant stars at a given luminosity (Salaris et al.~\cite{salaris02}). 
Below log T = 3.75 we use the data from
Ferguson et al.~(\cite{ferguson05}) (publicly available at http://webs.wichita.edu/physics/opacity/).
In the higher temperature regime we use the OPAL opacities
computed from the Lawrence Livermore National
Laboratory web interface (http://physci.llnl.gov/Research/OPAL/new.html).
Both opacity sets correspond to the solar metal repartition 
advocated by Asplund et al.~(\cite{aspl05}).
The electron conductive opacities that are only significant 
in the degenerate helium cores of red giants 
follow the Itoh et al.~(\cite{itoh83}) prescription.
We model stars well below the RGB tip
and using the newer prescription of conductive opacities
by Cassisi et al.~(\cite{cassisi07}) would likely 
leave the evolution tracks unchanged as noticed by these authors.

{\it ii) The convection}:
Two simplified local treatments of convection are investigated:
the mixing length theory and the full spectrum of turbulence
model. We chose a prescription of the MLT
very similar to that of B\"{o}hm-Vitense~(\cite{bohm58})
and whose exact description is given in the appendix of 
Piau et al.~(\cite{piau05}). The formulation of the FST is 
that of Canuto, Goldman \& Mazzitelli~(\cite{cgm96}) and
referred as CGM hereafter. The detailed equations are provided
in the second appendix to this article. For the CGM approach, we adopt a characteristic
length scale $\rm \Lambda=\alpha_{cgm} H_p$ in the equations 
-where $\rm H_p$ is the pressure scale height-.
The CGM version we use 
is a simplified version rather than the true version of the CGM
where $\rm \Lambda$ is the distance to the
convection/radiation regime boundary.
Finally we have two length scale parameters
for the two treatments of convection:
$\rm \alpha_{mlt}$ and $\rm \alpha_{cgm}$.
In the occurence of convective
cores, i.e. for stars more massive than $\rm 1.2 M_{\odot}$,
we assume convective overshooting ranging from $\rm 0.1 H_p$ to $\rm 0.2 H_p$
(Claret~\cite{claret07}).

{\it iii) The atmosphere modelling}:
We define the atmosphere as the region where 
the Rosseland optical depth is less than $\rm \tau_b=20$.
The outer boundary conditions to the internal structure 
are taken at this depth where the diffusion equation 
becomes valid (Morel et al.~\cite{morel94}).
In late type stars, the convection straddles 
such an atmosphere/interior boundary and because the MLT and the CGM model
predict different temperature gradients, it is important to use
the same treatment of convection above and below
this limit (Montalban et al.~\cite{montal01}, 
Montalban et al.~\cite{montal04}). Besides this, molecular
lines have a significant impact on atmosphere structures
when $\rm T_{eff} < 5000~K$. For these reasons we introduce 
two series of non grey atmosphere models as outer boundary conditions.
The first series of temperature-optical depth 
relations ($\rm T(\tau)^4=T_{eff}^4 f_{grid}(\tau)$) is computed
with the PHOENIX/1D atmosphere code where the convection is handled 
using the MLT. The second series of temperature-optical depth relations is computed
with the ATLAS12 atmosphere code (Castelli~\cite{castelli05}). We modified
ATLAS12 in order to use the CGM version of the FST model of convection instead of 
the MLT. The change is directly inspired from the implementation 
of the FST to the Atlas9 code by Kupka~(\cite{kupka96}). 
The characteristic length scales adopted for the 
convection in both atmospheric sets is $\rm 0.5H_p$
as suggested by the solar and the cool dwarfs Balmer lines 
(Samadi et al.~\cite{sam06}; Gardiner et al.~\cite{gardiner99}).
Using the same convection model in the atmosphere and interior
is necessary for consistency. In the proper treatment of
the CGM, the convection length scale $\rm \Lambda$ is the distance to the
convection/radiation regime boundary. This 
mimics the increase of the convective efficiency 
with depth (Heiter et al.~\cite{hei02}) and the 
smooth change of the thermal gradient associated.
However this treatment is not handled by our code
where $\rm \Lambda=\alpha_{cgm}H_p$. In this framework,
it is impossible to fit the solar radius when 
using in the interior the (small) atmosphere 
$\rm \alpha_{cgm}$ required by the Balmer lines. 
The sudden change in $\rm \alpha_{cgm}$ (or $\rm \alpha_{mlt}$)
from the atmosphere to the interior should 
induce a discontinuity in the thermal gradient
at the limit between the interior and the atmosphere.
In order to smooth it out we perform a linear interpolation 
with optical depth on the temperature gradients:
provided $\rm \nabla_a$ and $\rm \nabla_i$ are the temperature gradients 
($\nabla = \frac{dlnT}{dlnP}$) computed using respectively the 
atmosphere and interior formalisms, we compute the gradient $\rm \nabla$ at the 
the optical depth $\rm \tau < \tau_b$ following: 
$$\rm \nabla = (1-x(\tau))\nabla_a + x(\tau) \nabla_i$$
where $\rm x(\tau)=max(0,(\tau-1)/(\tau_b-1))$.
Thus the atmospheric gradient is taken where $\rm \tau < 1$, the
interior gradient where $\rm \tau > \tau_b$ and and linear combination of
them where $\rm 1 < \tau < \tau_b$.

Both atmosphere grids were computed with the solar Asplund et al.~(\cite{aspl05}) 
composition. In both of them there is a model every
100 K on the [4000,6400]K effective temperature range and every 
0.5 on the [0.5,5.5] range of decimal logarithm of surface 
gravity (log g). For every stellar model 
we perform a linear interpolation in $\rm T_{eff}$ and log g using the four
closest neighbours on the grid in order to get the $\rm f(\tau)$ theoretical 
relations. Together with $\rm T_{eff}$ this relation 
provides the temperature atmosphere profile: $\rm T(\tau)^4=T_{eff}^4 f(\tau)$.
For the most massive stars modelled, $\rm T_{eff} > 6400~K$ on the main sequence and the
$\rm f_{grid}(\tau)$ function is not available. In this case we use
the $\rm f_{grid}(\tau)$ at 6400 K and for the closest log g.

{\it iv) The equation of state (hereafter EOS)}:
We use the OPAL 2005 EOS tables for Population I stars. Although the
effect of the EOS is somewhat less important than the effects
of opacities or atmosphere boundary conditions (Salaris et al.~\cite{salaris02})
it still determines the temperature gradient in the convection zone 
and therefore affects the radius.

Let us make a final comment on the composition of the models 
before moving to the next section. 
The atmosphere $\rm T(\tau)$ relations are especially computed
for the Asplund et al.~(\cite{aspl05}) composition and metal 
repartition. The EOS and opacities both in low and high temperature regimes assume the
same metal repartition. However we do not only consider the
solar composition. The spectroscopic analyses of the red giants
of our sample suggest them to have on average a slightly lower metallicity 
than the Sun and small but non negligible composition differences
(\S \ref{sec52}). The total metal mass fraction 
Z and helium mass fraction Y are changed to explore the impact
of composition.

%______________________________________________________________ PARTIE OBSERVATIONS

\section{The observations}\label{sec3}

\begin{figure}
\includegraphics[width=\hsize]{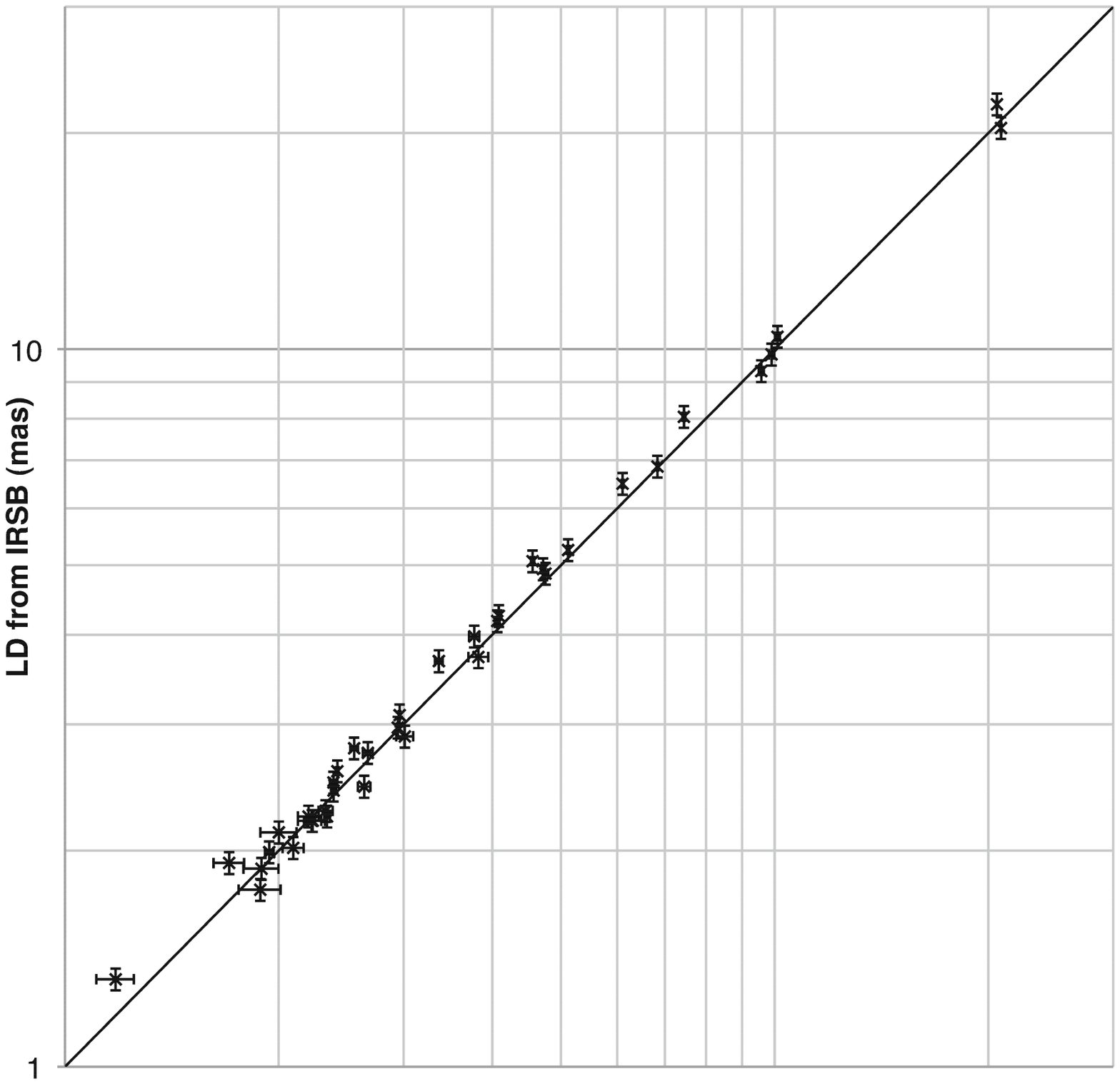}
\includegraphics[width=\hsize]{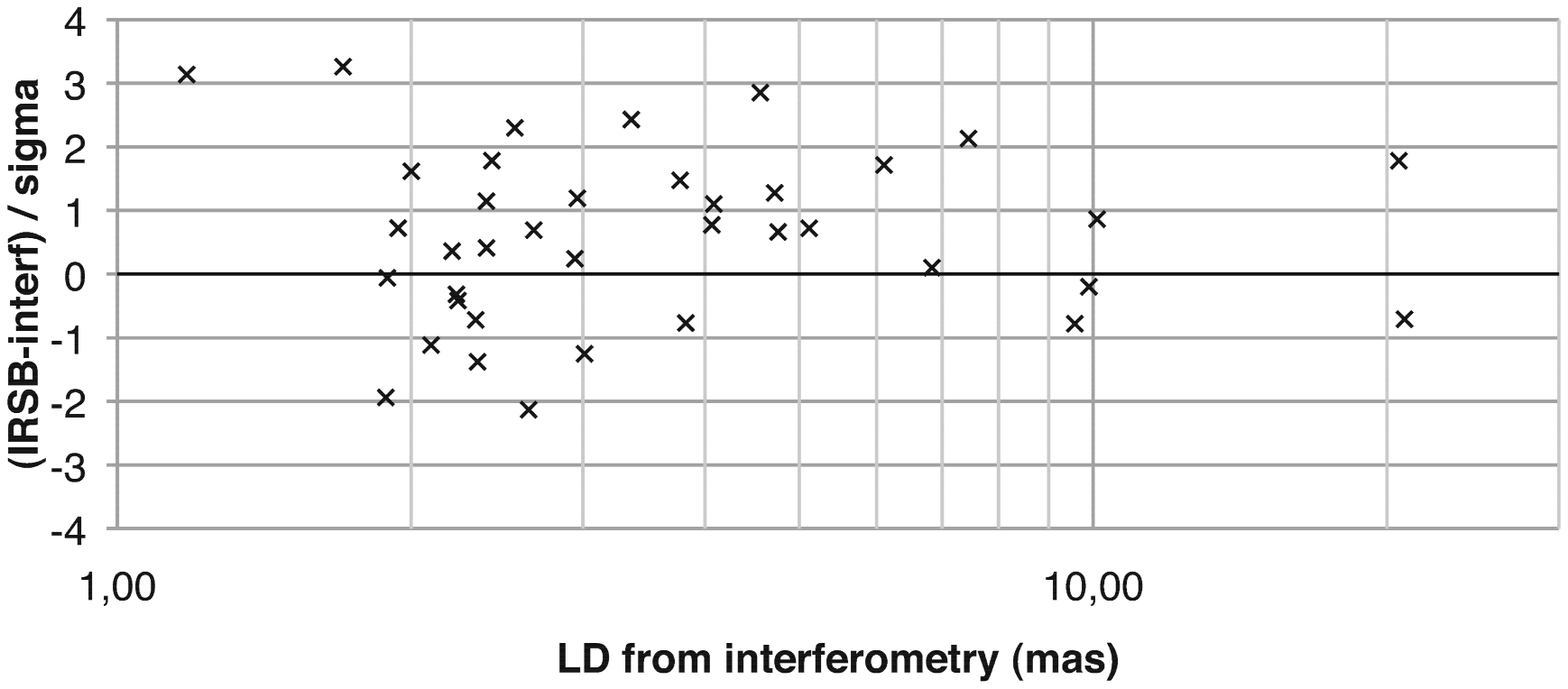}
\caption{Comparison of the limb darkened angular diameters of the stars of our sample from direct interferometric measurements (horizontal axis) and photometric estimates based on the $F_K(V-K)$ infrared surface brightness-color relation (IRSB) from Kervella et al.~(\cite{kervella04a}). The lower plot shows the residuals as a function of the measured angular diameter. \label{Interf_IRSB}} 
\end{figure}

The selection of our sample of stars was done in two steps. We first queried the CHARM2 catalogue (Richichi et al.~\cite{richichi05}) to obtain all direct measurements of giant and subgiant angular diameters up to 2004, with effective temperatures in the range 5000 to 5500\,K. The giants from Hutter et al.'s~(\cite{hutter89}) Mark III survey  were removed from the sample, as most of these measurements present strong biases due to calibration uncertainties (see their Sect. IV for details). We then searched the literature for more recent observations, and added the measurement of $\gamma$\,Sge by Wittkowski et al.~(\cite{wittkowski06}), $\delta$\,Eri and $\xi$\,Hya by Th\'evenin et al.~(\cite{thevenin05}) and the recent high accuracy CHARA/FLUOR measurements of $\epsilon$\,Oph (Mazumdar et al.~\cite{mazumdar09}) and $\eta$\,Ser (M\'erand et al.~\cite{merand09}). When several independent angular diameter measurements were available for the same star, we combined them into a single value taking into account their original error bars and the consistency of the different measurements. The conversion of uniform disk angular diameters to limb darkened values was done using linear limb darkening coefficients by Claret~(\cite{claret95}), that are based on stellar atmosphere models by Kurucz (Kurucz~\cite{kur93}). We are aware that the plane parallel ATLAS9 limb darkening coefficients by Claret~(\cite{claret95}) are not optimal for cool, giant stars with low effective gravities, but the difference with PHOENIX models is very small for the selected sample. This difference amounts typically to a few per mille of the stellar size. Choosing the Claret~(\cite{claret95}) values has the advantage of preserving the internal consistency of the sample, as the ATLAS9 models are the standard for the quoted interferometric measurements. A comparison of the measured angular diameters with predictions from the $F_K(V-K)$ surface brightness-color relations calibrated by Kervella et al.~(\cite{kervella04a}) on Cepheid supergiants is presented in Fig.~\ref{Interf_IRSB}. The agreement is satisfactory within the uncertainties, with no systematic bias at a level of 2\%. It is interesting to remark that these relations, established for supergiants, are mostly identical to the relations calibrated by Kervella et al.~(\cite{kervella04b}) using dwarf and subgiants. This is an indication that the infrared surface brightness-color relations are universal for all classes of F-K stars.

The spectroscopic effective temperature $\rm T_{eff}$, effective gravity $\log g$ and metallicity [Fe/H] of each star of our sample were taken from Cayrel de Strobel et al.'s~(\cite{cayrel01}) catalogue, except for $\gamma$\,Sge and $\xi$\,Hya for which we used the $\rm T_{eff}$ values determined in the articles reporting their interferometric angular diameters. These estimates were originally obtained by McWilliam~(\cite{mcwilliam90}), Th\'evenin \& Idiart~(\cite{thevenin99}) and Mallik~(\cite{mallik98}). Similarly the $\rm T_{eff}$, metallicity and luminosity of $\epsilon$\,Oph are adapted from De Ridder et al.~(\cite{der06}) (also in agreement with Mazumdar et al.~(\cite{mazumdar09})). The bolometric luminosity was estimated using the $K$ band photometry and the corresponding bolometric corrections $BC(K)$ by Houdashelt et al.~(\cite{houdashelt00}). For $\gamma$\,Sge, we extrapolated the $BC(K)$ value for $T_{\rm eff} = 3805$\,K (Wittkowski et al.~\cite{wittkowski06}). For $\epsilon$\,Oph and $\delta$\,Eri, we adopted the $\rm T_{eff}$ metallicity and luminosity from De Ridder et al.~(\cite{der06}) and Thevenin et al.~(\cite{thevenin05}) respectively. To derive the linear radius, we used the parallaxes from the Hipparcos catalogue~(\cite{esa97}). We kept only the stars for which the relative uncertainty on the linear radius is lower than 10\%. The distances to the selected stars range from 11 to 110\,pc. Thanks to this proximity, we neglected the interstellar reddening for the computation of the bolometric luminosity.

Our final sample contains 38 giant and subgiant stars with spectral types G5 to M0. It is interesting to remark that three stars of our sample have known asteroseismic oscillation frequencies: $\delta$\,Eri, $\xi$\,Hya, $\epsilon$\,Oph. This allowed an accurate determination of their masses through a combination of their radius and the large frequency spacing (Th\'evenin et al.~\cite{thevenin05}; Mazumdar et al.~\cite{mazumdar09}), that are estimated respectively to 1.215, 2.65 (no error bars mentioned by the authors) and $\rm 1.85 \pm 0.05\,M_\odot$.

\begin{landscape}
\begin{table}
\caption{Observational and physical parameters of our selected sample of giant and subgiant stars with interferometrically measured angular diameters.} 
\centering
\label{observations_table}
\begin{tabular}{lllllllllrrlcl}
\hline
\hline
Star & HD & Sp. Type & LD   & $\pi$   & Radius   & $T_{\rm eff}$ & $\log g$ & [Fe/H] & $m_V$ & $m_K$ & BC($K$) & $M_{\rm bol}$   & $L_{\rm bol}$   \\
 & & & (mas) & (mas) & ($R_\odot$) & (K) & & & & & & & ($L_\odot$) \\
\hline
$\mu$\,Her & \object{ HD 161797 } & G5IV & $ 1.94 \pm 0.03 $ & $ 119.05 \pm 0.62 $ & $ 1.75 \pm 0.03 $ & $ 5520 $ & $ 3.70 $ & $ 0.04 $ & $ 3.41 $ & $ 1.74 $ & $ 1.56 $ & $ 3.68 \pm 0.05 $ & $ 2.7 \pm 0.1 $ \\
$\delta$\,Eri & \object{ HD 23249 } & K0IV & $ 2.39 \pm 0.03 $ & $ 110.58 \pm 0.88 $ & $ 2.32 \pm 0.03 $ & $ 5055 $ & $ 3.79 $ & $ 0.13 $ & $ 3.51 $ & $ 1.40 $ & $ 1.84 $ & $ 3.46 \pm 0.05 $ & $ 3.3 \pm 0.2 $ \\
$\beta$\,Aql & \object{ HD 188512 } & G8IV & $ 2.23 \pm 0.10 $ & $ 72.95 \pm 0.83 $ & $ 3.28 \pm 0.15 $ & $ 5100 $ & $ 3.60 $ & $ -0.30 $ & $ 3.71 $ & $ 1.65 $ & $ 1.84 $ & $ 2.81 \pm 0.06 $ & $ 6.0 \pm 0.3 $ \\
$\eta$\,Cep & \object{ HD 198149 } & K0IV & $ 2.67 \pm 0.04 $ & $ 69.73 \pm 0.49 $ & $ 4.12 \pm 0.07 $ & $ 4950 $ & $ 3.41 $ & $ -0.32 $ & $ 3.41 $ & $ 1.22 $ & $ 1.84 $ & $ 2.28 \pm 0.05 $ & $ 9.7 \pm 0.5 $ \\
$\eta$\,Ser & \object{ HD 168723 } & K0III-IV & $ 2.94 \pm 0.01 $ & $ 53.93 \pm 0.18 $ & $ 5.87 \pm 0.03 $ & $ 4890 $ & $ 3.21 $ & $ -0.42 $ & $ 3.26 $ & $ 1.05 $ & $ 1.84 $ & $ 1.55 \pm 0.05 $ & $ 19 \pm 1 $ \\
46\,Lmi & \object{ HD 94264 } & K0III & $ 2.56 \pm 0.03 $ & $ 33.40 \pm 0.78 $ & $ 8.22 \pm 0.22 $ & $ 4670 $ & $ 2.96 $ & $ -0.20 $ & $ 3.83 $ & $ 1.30 $ & $ 2.00 $ & $ 0.91 \pm 0.07 $ & $ 34 \pm 2 $ \\
$\sigma$\,Gem   & \object{ HD 62044 } & K1III & $ 2.33 \pm 0.05 $ & $ 26.68 \pm 0.79 $ & $ 9.38 \pm 0.35 $ & $ 4500 $ & $ 2.40 $ & $ -0.02 $ & $ 4.28 $ & $ 1.74 $ & $ 2.16 $ & $ 1.03 \pm 0.08 $ & $ 31 \pm 2 $ \\
$\epsilon$\,Aql   & \object{ HD 176411} & K1III & $ 2.00 \pm 0.12 $ & $ 21.22 \pm 0.77 $ & $ 10.14 \pm 0.70 $ & $ 4760 $ & $ 2.91 $ & $ 0.00 $ & $ 4.03 $ & $ 1.79 $ & $ 2.00 $ & $ 0.42 \pm 0.09 $ & $ 54 \pm 5 $ \\
37\,Tau & \object{ HD 25604 } & K0III & $ 1.70 \pm 0.09 $ & $ 18.04 \pm 0.84 $ & $ 10.15 \pm 0.69 $ & $ 4700 $ & $ 2.77 $ & $ 0.01 $ & $ 4.35 $ & $ 2.03 $ & $ 2.00 $ & $ 0.31 \pm 0.11 $ & $ 60 \pm 6 $ \\
$\xi$\,Hya & \object{ HD 100407 } & G7III & $ 2.39 \pm 0.02 $ & $ 25.23 \pm 0.83 $ & $ 10.18 \pm 0.35 $ & $ 5037 $ & $ 2.83 $ & $ -0.04 $ & $ 3.54 $ & $ 1.45 $ & $ 1.84 $ & $ 0.30 \pm 0.09 $ & $ 60 \pm 5 $ \\
$\epsilon$\,Oph & \object{ HD 146791 } & G9.5IIIb & $ 2.96 \pm 0.01 $ & $ 30.62 \pm 0.20 $ & $ 10.39 \pm 0.07 $ & $ 4955 $ & $ 2.89 $ & $ -0.27 $ & $ 3.24 $ & $ 0.98 $ & $ 1.84 $ & $ 0.25 \pm 0.05 $ & $ 63 \pm 3 $ \\
$\delta$\,Ari   & \object{ HD 19787 } & K2III & $ 1.88 \pm 0.13 $ & $ 19.44 \pm 1.23 $ & $ 10.42 \pm 0.97 $ & $ 4810 $ & $ 2.93 $ & $ -0.03 $ & $ 4.35 $ & $ 2.17 $ & $ 2.00 $ & $ 0.61 \pm 0.14 $ & $ 45 \pm 6 $ \\
$\o$\,CrB   & \object{ HD 136512} & K0III & $ 1.18 \pm 0.07 $ & $ 11.90 \pm 0.74 $ & $ 10.64 \pm 0.93 $ & $ 4730 $ & $ 2.75 $ & $ -0.44 $ & $ 5.52 $ & $ 2.93 $ & $ 2.00 $ & $ 0.30 \pm 0.14 $ & $ 60 \pm 8 $ \\
$\epsilon$\,Cyg   & \object{ HD 197989} & K0III & $ 4.56 \pm 0.02 $ & $ 45.26 \pm 0.53 $ & $ 10.82 \pm 0.14 $ & $ 4730 $ & $ 2.89 $ & $ -0.27 $ & $ 2.50 $ & $ -0.01 $ & $ 2.00 $ & $ 0.26 \pm 0.06 $ & $ 62 \pm 3 $ \\
$\delta$\,Tau   & \object{ HD 27697 } & K0III & $ 2.20 \pm 0.07 $ & $ 21.29 \pm 0.93 $ & $ 11.12 \pm 0.61 $ & $ 4810 $ & $ 2.93 $ & $ -0.03 $ & $ 3.76 $ & $ 1.64 $ & $ 2.00 $ & $ 0.28 \pm 0.11 $ & $ 62 \pm 6 $ \\
39\,ari & \object{ HD 17361 } & K1.5III & $ 1.89 \pm 0.11 $ & $ 18.06 \pm 0.84 $ & $ 11.26 \pm 0.82 $ & $ 4600 $ & $ 2.85 $ & $ -0.02 $ & $ 4.51 $ & $ 2.10 $ & $ 2.16 $ & $ 0.54 \pm 0.11 $ & $ 48 \pm 5 $ \\
$\delta^1$\,Tau & \object{ HD 27697 } & K0III & $ 2.34 \pm 0.02 $ & $ 21.29 \pm 0.93 $ & $ 11.81 \pm 0.53 $ & $ 4030 $ & $ 1.83 $ & $ -0.29 $ & $ 3.76 $ & $ 1.64 $ & $ 2.51 $ & $ 0.79 \pm 0.11 $ & $ 39 \pm 4 $ \\
12\,Aql & \object{ HD 176678} & K1III & $ 2.42 \pm 0.01 $ & $ 21.95 \pm 0.92 $ & $ 11.85 \pm 0.50 $ & $ 4600 $ & $ 2.75 $ & $ -0.19 $ & $ 4.03 $ & $ 1.47 $ & $ 2.16 $ & $ 0.34 \pm 0.10 $ & $ 58 \pm 6 $ \\
$\delta$\,And   & \object{ HD 3627  } & K3III & $ 4.09 \pm 0.02 $ & $ 32.19 \pm 0.68 $ & $ 13.64 \pm 0.30 $ & $ 4360 $ & $ 2.37 $ & $ 0.04 $ & $ 3.28 $ & $ 0.47 $ & $ 2.16 $ & $ 0.17 \pm 0.07 $ & $ 68 \pm 4 $ \\
$\alpha$\,Ari   & \object{ HD 12929 } & K2III & $ 6.84 \pm 0.02 $ & $ 49.48 \pm 0.99 $ & $ 14.85 \pm 0.30 $ & $ 4480 $ & $ 2.57 $ & $ -0.25 $ & $ 2.00 $ & $ -0.63 $ & $ 2.16 $ & $ -0.15 \pm 0.07 $ & $ 91 \pm 6 $ \\
$\rho$\,Boo   & \object{ HD 127665} & K3III & $ 3.82 \pm 0.12 $ & $ 21.92 \pm 0.81 $ & $ 18.75 \pm 0.92 $ & $ 4260 $ & $ 2.22 $ & $ -0.17 $ & $ 3.58 $ & $ 0.76 $ & $ 2.33 $ & $ -0.21 \pm 0.09 $ & $ 96 \pm 9 $ \\
$\psi$\,Uma   & \object{ HD 96833 } & K1III & $ 4.07 \pm 0.03 $ & $ 22.21 \pm 0.68 $ & $ 19.68 \pm 0.62 $ & $ 4550 $ & $ 2.53 $ & $ -0.13 $ & $ 3.01 $ & $ 0.43 $ & $ 2.16 $ & $ -0.68 \pm 0.08 $ & $ 148 \pm 12 $ \\
$\upsilon$\,Per   & \object{ HD 9927  } & K3III & $ 3.77 \pm 0.06 $ & $ 18.76 \pm 0.74 $ & $ 21.61 \pm 0.93 $ & $ 4380 $ & $ 2.34 $ & $ 0.00 $ & $ 3.57 $ & $ 0.65 $ & $ 2.16 $ & $ -0.83 \pm 0.10 $ & $ 170 \pm 16 $ \\
24\,Per & \object{ HD 18449 } & K2III & $ 2.10 \pm 0.07 $ & $ 9.31 \pm 0.78 $ & $ 24.21 \pm 2.19 $ & $ 4340 $ & $ 2.37 $ & $ -0.19 $ & $ 4.95 $ & $ 2.10 $ & $ 2.33 $ & $ -0.73 \pm 0.18 $ & $ 155 \pm 28 $ \\
$\alpha$\,Boo   & \object{ HD 124897} & K1.5III & $ 20.84 \pm 0.03 $ & $ 88.85 \pm 0.74 $ & $ 25.21 \pm 0.21 $ & $ 4345 $ & $ 2.05 $ & $ -0.37 $ & $ -0.04 $ & $ -2.91 $ & $ 2.16 $ & $ -1.01 \pm 0.05 $ & $ 201 \pm 10 $ \\
91\,Psc & \object{ HD 8126  } & K5III & $ 2.23 \pm 0.06 $ & $ 9.49 \pm 0.82 $ & $ 25.30 \pm 2.29 $ & $ 4090 $ & $ 1.93 $ & $ -0.17 $ & $ 5.24 $ & $ 2.03 $ & $ 2.51 $ & $ -0.58 \pm 0.19 $ & $ 135 \pm 25 $ \\
39\,Cyg & \object{ HD 194317} & K3III & $ 3.01 \pm 0.08 $ & $ 12.77 \pm 0.62 $ & $ 25.35 \pm 1.41 $ & $ 4230 $ & $ 2.16 $ & $ -0.17 $ & $ 4.44 $ & $ 1.39 $ & $ 2.33 $ & $ -0.75 \pm 0.11 $ & $ 159 \pm 18 $ \\
11\,Lac & \object{ HD 214868} & K2III & $ 2.64 \pm 0.05 $ & $ 10.81 \pm 0.56 $ & $ 26.24 \pm 1.46 $ & $ 4440 $ & $ 2.32 $ & $ -0.25 $ & $ 4.51 $ & $ 1.67 $ & $ 2.16 $ & $ -1.00 \pm 0.12 $ & $ 200 \pm 23 $ \\
31\,Leo & \object{ HD 87837 } & K4III & $ 3.36 \pm 0.04 $ & $ 11.89 \pm 0.72 $ & $ 30.40 \pm 1.88 $ & $ 4040 $ & $ 1.81 $ & $ -0.02 $ & $ 4.38 $ & $ 0.98 $ & $ 2.51 $ & $ -1.14 \pm 0.14 $ & $ 227 \pm 31 $ \\
$\upsilon$\,Boo   & \object{ HD 120477} & K5.5III & $ 4.75 \pm 0.04 $ & $ 13.29 \pm 0.81 $ & $ 38.44 \pm 2.37 $ & $ 3890 $ & $ 1.55 $ & $ -0.23 $ & $ 4.05 $ & $ 0.44 $ & $ 2.51 $ & $ -1.44 \pm 0.14 $ & $ 299 \pm 40 $ \\
$\beta$\,Umi   & \object{ HD 131873} & K4III & $ 10.09 \pm 0.08 $ & $ 25.79 \pm 0.52 $ & $ 42.06 \pm 0.91 $ & $ 4030 $ & $ 1.83 $ & $ -0.29 $ & $ 2.08 $ & $ -1.29 $ & $ 2.51 $ & $ -1.73 \pm 0.07 $ & $ 390 \pm 25 $ \\
$\alpha$\,Tau   & \object{ HD 29139 } & K5III & $ 20.57 \pm 0.02 $ & $ 50.09 \pm 0.95 $ & $ 44.13 \pm 0.84 $ & $ 3910 $ & $ 1.59 $ & $ -0.34 $ & $ 0.86 $ & $ -2.81 $ & $ 2.51 $ & $ -2.04 \pm 0.06 $ & $ 518 \pm 32 $ \\
$\gamma$\,Dra   & \object{ HD 164058} & K5III & $ 9.90 \pm 0.09 $ & $ 22.10 \pm 0.46 $ & $ 48.15 \pm 1.09 $ & $ 3930 $ & $ 1.55 $ & $ -0.14 $ & $ 2.23 $ & $ -1.16 $ & $ 2.51 $ & $ -1.93 \pm 0.07 $ & $ 471 \pm 30 $ \\
$\beta$\,Cnc   & \object{ HD 69267 } & K4III & $ 5.12 \pm 0.02 $ & $ 11.23 \pm 0.97 $ & $ 48.96 \pm 4.23 $ & $ 4010 $ & $ 1.71 $ & $ -0.24 $ & $ 3.54 $ & $ 0.19 $ & $ 2.51 $ & $ -2.05 \pm 0.19 $ & $ 526 \pm 99 $ \\
$\alpha$\,Lyn   & \object{ HD 80493 } & K7III & $ 7.45 \pm 0.04 $ & $ 14.69 \pm 0.81 $ & $ 54.50 \pm 3.02 $ & $ 3880 $ & $ 1.51 $ & $ -0.26 $ & $ 3.13 $ & $ -0.61 $ & $ 2.51 $ & $ -2.32 \pm 0.13 $ & $ 673 \pm 83 $ \\
$\gamma$\,Sge & \object{ HD 189319 } & M0III & $ 6.10 \pm 0.02 $ & $ 11.90 \pm 0.71 $ & $ 55.13 \pm 3.29 $ & $ 3805 $ & $ 1.55 $ & $ -0.14 $ & $ 3.53 $ & $ -0.16 $ & $ 2.66 $ & $ -2.13 \pm 0.14 $ & $ 562 \pm 75 $ \\
$\alpha$\,Hya   & \object{ HD 81797 } & K3II-III & $ 9.58 \pm 0.08 $ & $ 18.40 \pm 0.78 $ & $ 55.93 \pm 2.41 $ & $ 4120 $ & $ 1.77 $ & $ -0.12 $ & $ 2.00 $ & $ -1.13 $ & $ 2.33 $ & $ -2.48 \pm 0.10 $ & $ 780 \pm 78 $ \\
$\nu$\,UMa   & \object{ HD 98262 } & K3III & $ 4.72 \pm 0.03 $ & $ 8.88 \pm 0.64 $ & $ 57.07 \pm 4.13 $ & $ 4070 $ & $ 1.89 $ & $ -0.04 $ & $ 3.50 $ & $ 0.28 $ & $ 2.51 $ & $ -2.47 \pm 0.16 $ & $ 775 \pm 122 $ \\
\hline

\end{tabular}
\begin{list}{}{}
\item[$^{\mathrm{a}}$] $\theta_{\rm LD}$ is the limb darkened angular diameter measured by interferometry, taken from the CHARM2 catalogue compiled by Richichi et al.~(\cite{richichi05}), except for $\gamma$\,Sge (Wittkowski et al.~\cite{wittkowski06}), $\delta$\,Eri and $\xi$\,Hya (Th\'evenin et al.~\cite{thevenin05}), $\epsilon$\,Oph (Mazumdar et al.~\cite{mazumdar09}) and $\eta$\,Ser (M\'erand et al.~\cite{merand09}).
\item[$^{\mathrm{b}}$] The parallax $\pi$ is taken from the original \emph{Hipparcos} catalogue (ESA~\cite{esa97}).
\item[$^{\mathrm{c}}$] $T_{\rm eff}$, $\log g$ and [Fe/H] are taken from Cayrel de Strobel et al.~(\cite{cayrel01}), except for $\gamma$\,Sge, $\delta$\,Eri, $\epsilon$\,Oph and $\xi$\,Hya (see text).
\item[$^{\mathrm{d}}$] $m_V$ and $m_K$ were taken from SIMBAD.
\item[$^{\mathrm{e}}$] The uncertainty on the $K$ band bolometric correction $BC(K)$ (taken from Houdashelt et al.~\cite{houdashelt00}, except for $\gamma$\,Sge) is taken uniformly equal to 0.05.
\item[$^{\mathrm{f}}$] The absolute bolometric magnitude of the Sun is $M_{\rm bol}(\odot)=4.75$.
\end{list}
\end{table}
\end{landscape}

%______________________________________________________________ PARTIE OBSERVATIONS

%On defini l'echantillon des 50 meilleures etoiles : les criteres
%sont la precision sur le rayon absolu (precision sur la mesure angulaire + precision
%sur la distance) et la metallicite qui doit etre la plus proche de la
%metallicite solaire.

\section{Convection length scales from the Sun}\label{sec4}

The convection length scales can be calibrated using the Sun,
as is customarily done. We assume $\rm L_{\odot}=3.846\,10^{33} erg.s^{-1}$ and 
$\rm R_{\odot}=6.9599\,10^{10} cm$ and 
start the solar evolution on the zero age main sequence.
The calibration in luminosity, radius and metal to hydrogen 
ratio $\rm \frac{Z_{surf}}{X_{surf}}$
are achieved to better than $10^{-4}$ 
at the age of 4.6 Gyr for both MLT and CGM convection
prescriptions. We consider the solar abundances and metal 
repartition of Asplund et al. (2005) therefore 
$\rm \frac{Z_{surf}}{X_{surf}}=1.65\,10^{-2}$.
The microscopic diffusion is accounted for
following Proffitt and Michaud~(\cite{proffitt93})
It induces a decrease in helium and metal surface fractions in the 
course of the solar main sequence.

In the MLT framework we obtain $\rm \alpha_{mlt\odot}$=1.98. 
The initial composition of the calibrated
solar model is $\rm X_{o}=0.7284$, $\rm Y_{o}=0.2578$
and $\rm Z_{o}=1.38\,10^{-2}$
and its solar age surface composition is 
$\rm X_{surf}=0.7592$ and $\rm Y_{surf}=0.2282$.
The $\rm \alpha_{mlt\odot}$ is compatible with the
recent estimate of VandenBerg et al.~(\cite{vdb08}) also using non grey atmospheres
as boundary conditions. Using the MARCS atmosphere models (Gustafsson et al.~\cite{gus03})
the latter authors find $\rm \alpha_{mlt\odot}$=2.01. Samadi et al.~(\cite{sam06}) 
using the Kurucz atmosphere models (Kurucz~\cite{kur93}, 
see also Heiter et al.~\cite{hei02}) find $\rm \alpha_{mlt\odot}$=2.53.
This larger value is likely due to the fact that
Samadi et al. start interpolating between the atmosphere and interior gradients
at the optical depth $\rm \tau=4$ whereas we start 
at $\rm \tau=1$ (see section \ref{sec2}).
The overadiabaticity induced by the atmosphere $\rm \alpha=0.5$ extends over 
a larger region in their models and has to be compensated 
by a more efficient convection deeper.
We recall that $\rm \alpha_{mlt\odot}$ depends on the
assumed solar composition. For the Grevesse \& Sauval~(\cite{gs98})
composition ($\rm \frac{Z_{surf}}{X_{surf}}=2.3\,10^{-2}$) 
calibrated Sun, we find $\rm \alpha_{mlt\odot}$=1.67.

In the CGM model framework we obtain $\rm \alpha_{cgm\odot}$=0.77. 
The initial composition of the calibrated CGM 
solar model is nearly the same as for the calibrated MLT
solar model. $\rm Z_{o}=1.38\,10^{-2}$ 
is identical while the hydrogen and helium fractions
very slightly differ from the MLT case: $\rm X_{o}=0.7280$ and $\rm Y_{o}=0.2582$.
So is the current surface composition with
$\rm X_{surf}=0.7587$ and $\rm Y_{surf}=0.22865$.
Our $\rm \alpha_{cgm\odot}$ is in close agreement 
with the values of authors who used the 
same formulation of the CGM. Bernkopf~(\cite{bernkopf98}) finds $\rm \alpha_{cgm\odot}=0.82$
and Samadi et al.~(\cite{sam06}) find $\rm \alpha_{cgm\odot}=0.78$.

Theoretical atmosphere models probably account properly for the
differential changes in the outer stellar regions with $\rm T_{eff}$ and surface
gravity. Yet, in the solar case, empirical atmosphere
relations seem to reproduce the limb darkening
better than theoretical ones (Blackwell et al.~\cite{blckwl95}). Following VandenBerg et al.~(\cite{vdb08}) 
we combined the advantages of theoretical
and empirical $\rm T(\tau)$ relations. To that extent the $\rm T(\tau)$ we used is 
the interpolated theoretical ATLAS12 relations $\rm T_{theory}(\tau)$ 
rescaled by the empirical $\rm T_{HM74}(\tau)$ relation for the solar 
atmosphere from Holweger \& Mueller~(\cite{hm74}) as suggested by VandenBerg et al.~(\cite{vdb08}):
\begin{equation}
\rm T(\tau)= T_{theory}(\tau) + \frac{T_{eff}}{T_{eff,\odot}} [ T_{HM74}(\tau)- T_{theory,\odot}(\tau)]
\label{eq1}
\end{equation}
If we introduce such semi empirical atmosphere 
models the solar calibration leads to a slightly larger 
$\rm \alpha_{cgm\odot}$=0.80 for the CGM 
prescription. The initial
and final solar surface composition change negligibly.
We will discuss the influence of such a semi empirical 
atmosphere model in section \ref{sec53} below.

It is worth giving a word of caution on solar calibrations of convective
length scales. They are as reliable as are the solar models.
The dynamics of the Sun radiation zone is largely uncertain
as suggested by helioseismology (Turck-Chi\`eze et al.~\cite{turck04}).
It probably involves angular momentum
transport through internal waves (Charbonnel \& Talon~\cite{charb05};
Mathis et al.~\cite{mathis08}).
Moreover the mixing of the radiative interior associated with 
rotation (as in the tachocline) interacts with diffusion 
along the evolution (Brun, Turck-Chi\`eze \& Zahn~\cite{brun99}). 
This changes the actual surface composition.
In this respect, it is interesting to remark that
the calibration of the solar models using its
recent composition determinations have led to
helium content of the convection zone that systematically
underestimate the measurement made thanks to
seismology (Basu \& Antia~\cite{basu95}).

%We note that at optical depth $\rm \tau=2/3$
%the temperature is $T$ and the pressure.
%There are two critical factors in the determination of solar
%mixing length parameter: the atmosphere boundary condition
%to the internal structure and the low temperature opacity (cf \ref{sec2}).

\section{Red Giant branch calibrations}\label{sec5}

\subsection{The cool edge of the RGB}\label{sec51}

\begin{figure}[Ht]
\centering
\includegraphics[angle=90,width=8cm]{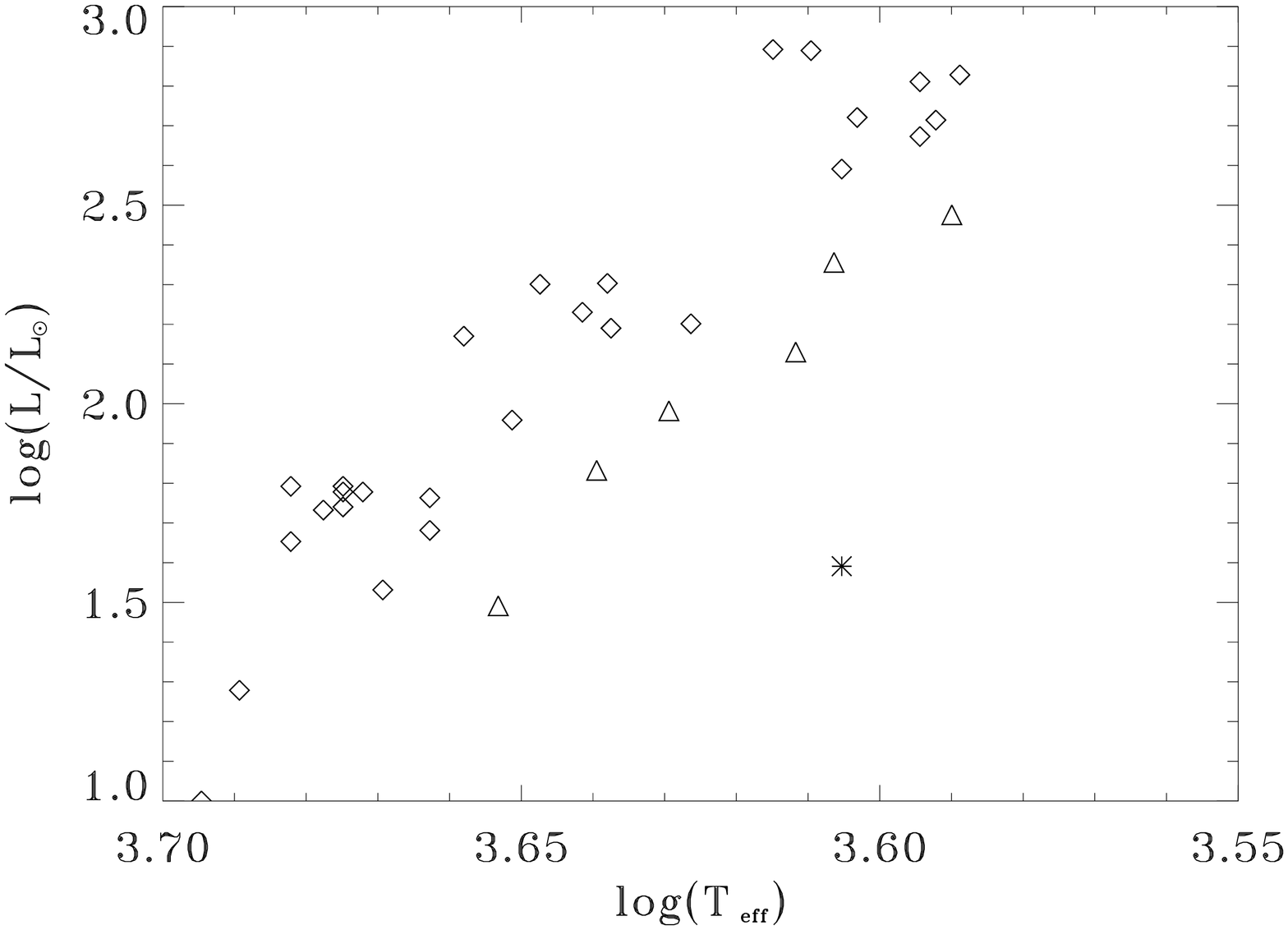}
\caption{Distribution of the current stellar sample in the HR diagram. Triangles: 
lower envelope of the RGB, see text for the names of the six corresponding 
objects. Diamonds: the other giants. Star: $\delta^1$Tau. 
\label{fig0a}} % hrgb1.pro 
\centering
\includegraphics[angle=90,width=8cm]{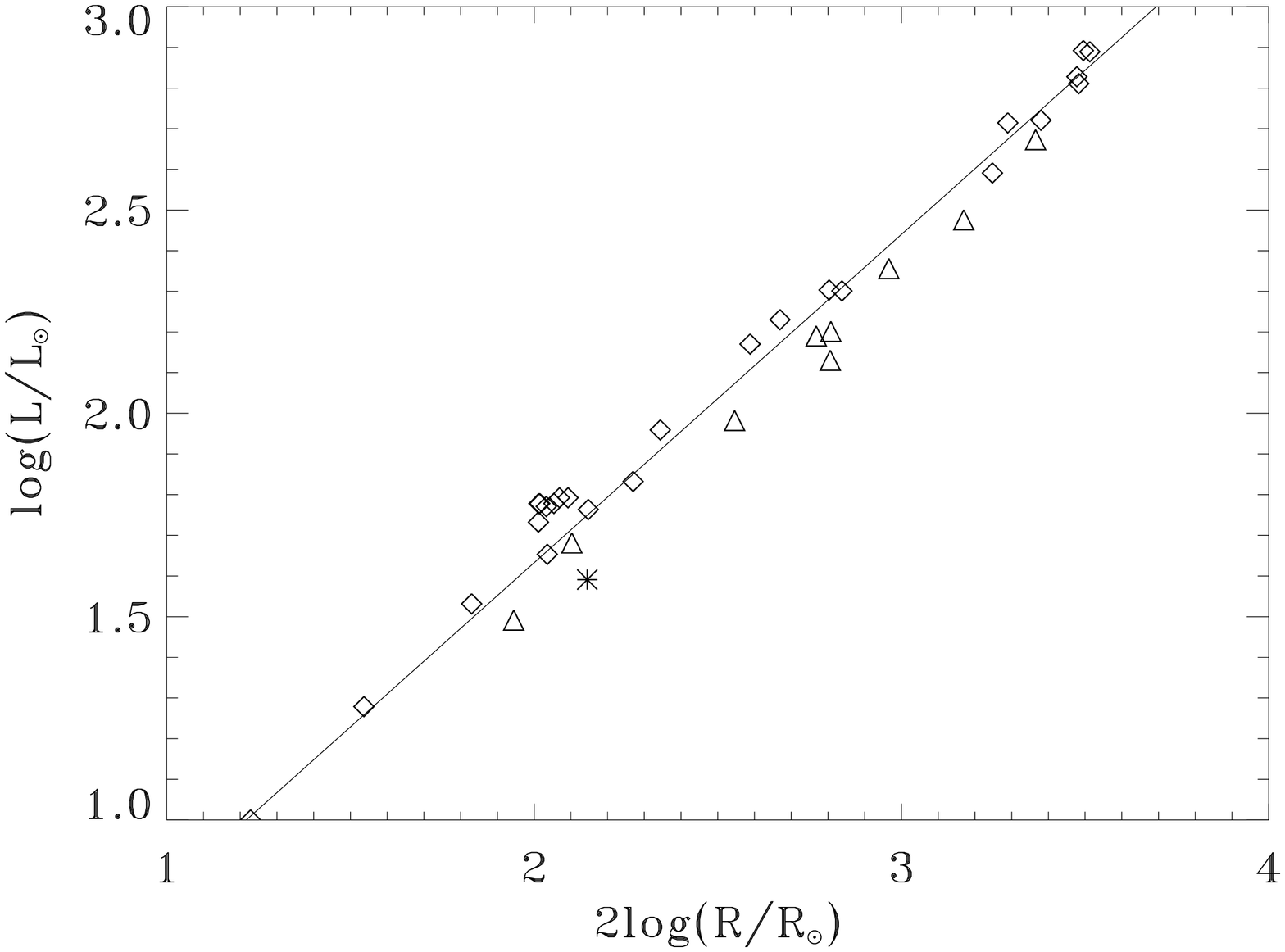}
\caption{Distribution of the current stellar sample in the log luminosity vs 
log squared radius diagram. The solid line is a linear fit to the 
sample by minimization of $\chi^2$. Triangles: group II, the nine giants
having the largest radius with respect to the corresponding 
fitted radius of the sample. Diamonds: the other giants. 
Star: $\delta^1$Tau. \label{fig0b}} % hrgb1.pro : Chi2 en R**2=0.342 
\end{figure}

Depending on its mass, a red giant will not be at the 
same effective temperature for a given luminosity. 
In the case of open or globular clusters all the
objects on the RGB approximately have the same mass
and the RGB is well defined.
This is because clusters gather stars of the same age.
No such assumption can be made for a sample that is representative of
field stars in the solar neighbourhood: stars of 
different ages (and thus masses) are presumably 
on the RGB we consider. 

It is however possible to set a lower limit 
on these masses. The limit is given by the age of the local Galactic disk
and the evolutionary time scale of its low mass stars:

{\it Age:} For objects in the slightly subsolar metallicity 
range ($\rm -0.25 <[Fe/H] < -0.14$) Liu \& Chaboyer~(\cite{liu00})
suggest a maximum age of $\rm 11.7\pm 1.9$ Gyr. 
In this study we mainly consider models that have
reached $\rm 10^3 L_{\odot}$ on the RGB by that age
as our RGB stars exhibits $\rm [Fe/H]=-0.17$ on average. 
We can further set upper limits to the local Galactic disk age: 
it is probably younger than the oldest Galaxy globular clusters whose current 
age estimate is 12.6 Gyr (Krauss \& Chaboyer~\cite{kra03})
and certainly younger than the
Universe $\rm 13.7 \pm 0.1$ Gyr (Komatsu et al.~\cite{komatsu09}).

{\it Initial composition:} The average metallicity of 
the 34 giants and 4 subgiants in Table\ref{observations_table} is $\rm [Fe/H]\sim-0.17$. 
As is well known the first dredge-up
occuring near the base of the RGB erases most of earlier diffusion effects
so that the actual surface metal fraction is also
the initial one.
We set the initial helium content at $\rm Y=0.2582$ i.e. the amount 
of our solar calibrated models. This choice could be criticized
as the stellar nucleosynthesis simultaneously increases 
the metal and helium fractions and both of them appear (loosely) correlated 
(Fernandes et al.~\cite{fern98}). Furthermore the estimate of the initial solar
content helium is still a matter of debate 
as it is affected by the dynamical phenomena in 
the solar interior (cf final caveat of \S \ref{sec4}). However a 
discussion on helium would be irrelevant 
here because even a substantial change in its initial fraction 
has a negligible impact on the position of the RGB as we will see shortly. 
Metals on the opposite have a very strong influence
on the effective temperatures and radii of RGB stars.
Thus in the following we repeat most of the 
calculations for the metallicities $\rm [Fe/H]=-0.17$
and $\rm [Fe/H]=0$. The fraction $\rm Z=1.20\,10^{-2}$
of the solar metallicity models very slightly differs
from our solar calibrated models $\rm Z=1.26\,10^{-2}$ 
because of the helium fraction difference between both sets.
In \ref{sec54} we performed additional calculations for
$\rm [Fe/H]=-0.34$ and built models with compositions dedicated to
the few stars of the sample having seismic data. The main
compositions investigated are given in details in Table \ref{Tabcomp1}.

\begin{table}[Ht]
  \caption[]{The three main different metallicity models.}
   \centering
     \label{Tabcomp1}
 $$ 
     \begin{array}{lcccccc}
        \hline
        \noalign{\smallskip}
        \rm [Fe/H]     & \rm Metals          & \rm  Hydrogen         \\
        \rm            & \rm mass\,fraction  & \rm  mass\,fraction   \\

        \noalign{\smallskip}			       
        \hline					       
        \noalign{\smallskip}			       
           0.0         &      1.20\,10^{-2}         &  0.7297  \\
          -0.17        &      8.18\,10^{-3}         &  0.7336  \\
          -0.34        &      5.55\,10^{-3}         &  0.7362  \\
        \noalign{\smallskip}
        \hline
     \end{array}
 $$ 
\end{table}

Between $\sim$30 and  $\sim$300 $\rm L_{\odot}$,
six stars clearly define the lower envelope of the local
Galactic RGB in the classical HR diagram: $\sigma$ Gem, 
$\delta$ And, $\rho$ Boo, 91 Psc, 31 Leo and $\upsilon$ Boo.
Hereafter we refer to them as 'group I'.
Fig.~\ref{fig0a} displays them and 
also shows an object that is far off the 
general trend. This is $\delta^1$Tau and it will not be 
considered in the following analysis.
In the bolometric luminosity vs. stellar squared radius
diagram the lower envelope of the RGB is harder
to see: Fig.~\ref{fig0b}. We selected the giants 
having the largest radii at a given luminosity
in the following manner: first we performed
a linear fit to the whole set of data. Then we kept only
the stars with the ten largest differences in
radius to the fitted radius at similar luminosities:
$\sigma$ Gem, 39 Ari, $\rho$Boo, 24 Per, 91 Psc, 39 Cyg, 31 Leo,
$\upsilon$Boo, $\beta$UMi and $\gamma$ Dra.
As in the case of the classical HR diagram 
$\delta^1$Tau was excluded from the sample.
$\beta$UMi was also excluded for its metallicity 
is very low ($\rm [Fe/H]=-0.29$) and we refer to the nine remaining 
stars as 'group II'.
Out of six stars making the lower envelope of the RGB 
in the classical HR diagram five also belong to the group of the nine 
most expanded objects. One could 
suggest that for simplicity we rely on the coolest stars also in the 
subsequent analysis using the radii. 
We keep groups I and II instead. One purpose of
this work is to use data on effective temperatures
and radii independently: the radii are determined 
through interferometry while the effective temperatures 
are determined through spectroscopy. Within the groups, the stars can be further 
classified depending on the metallicity \footnote{We did not find error bars on metallicities 
of group I and II stars in the catalog of Cayrel de Strobel~(\cite{cayrel01}).
We just mention that the error bars, when present in this catalog,
are generally around 0.1 dex}. 
$\sigma$ Gem, $\delta$ And, 31 Leo, 39 Ari all are within 0.04 dex 
of the solar metallicity. All the other members of groups I and II 
are within 0.06 dex of $\rm [Fe/H]=-0.17$ (and 
within 0.03 dex of this value if $\upsilon$ Boo is not considered).
Let us now compare the data on these objects to the evolutionary tracks
of models having different compositions, masses
and convection treatment length scales, first in the case of the MLT and then in 
the case of the CGM model.

\subsection{The cool edge of the RGB and the MLT}\label{sec52}

\begin{figure}[Ht]
\centering
\includegraphics[angle=90,width=8cm]{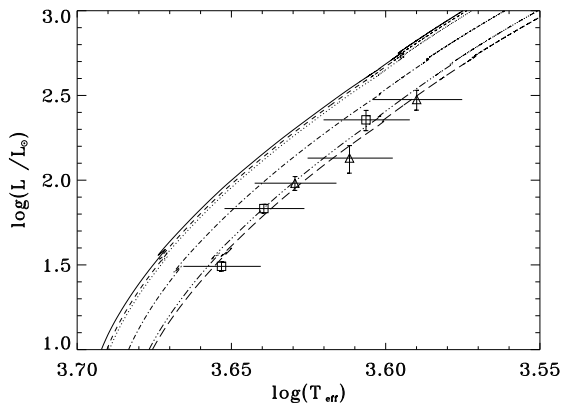}
\caption{RGBs in the HR diagram for models mlt1 to mlt6 of Table \ref{Tabmod1}.
The solid line is model mlt1, the dotted line 
is model mlt2, the dashed line is model mlt3, the dot-dashed line is 
model mlt4, the three-dot dashed line is model mlt5 and
the long dashed line is model mlt6. Models mlt2 and mlt3 show
mass and helium fraction change with respect to model mlt1.
They are hardly discernable from this model.
Model mlt4 shows the metallicity effect with respect to model mlt1.
Only models mlt5 and mlt6 with lower $\rm \alpha_{mlt}$ than 
the previous models fit the data satisfactorily.
The data are the six giants of 'group I' defined in the text and Fig.~\ref{fig0a}.
Square symbols are for objects within 0.04 dex of solar metallicity while 
triangles are for objects within 0.06 dex of $\rm [Fe/H]=-0.17$.
The error bars in $\rm log(T_{eff})$ are taken from McWilliam~(\cite{mcwilliam90}),
the error bars in log luminosity are adapted from Table 
\ref{observations_table}. \label{fig1a}} % hrgb1.pro 
\end{figure}
\begin{figure}[Ht]
\centering
\includegraphics[angle=90,width=8cm]{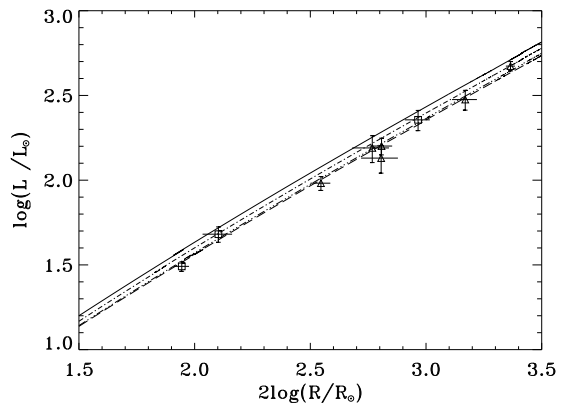}
\caption{RGBs in a bolometric luminosity vs. stellar squared radius
diagram for models mlt1, mlt4, mlt5 and mlt6 of Table \ref{Tabmod1}. 
The lines conventions 
are similar as in Fig.~\ref{fig1a}. Models mlt2 and mlt3
are not shown as they could hardly be distinguished from the track
of model mlt1.
The data are the nine giants of group II defined in text. Like in
Fig.~\ref{fig1a} square symbols
are for objects within 0.04 dex of solar metallicity, triangles 
for objects within 0.06 dex of $\rm [Fe/H]=-0.17$.
The best fits are obtained for the models mlt5 and mlt6. 
\label{fig1b}} % hrgb1.pro 
\end{figure}

Table \ref{Tabmod1} sums up the properties of the different 
models built using the MLT while Fig.~\ref{fig1a} and \ref{fig1b}
display their RGB evolutionary tracks.
In the Table, column 1 is the models mass, column 2 the metallicity, 
column 3 the helium mass fraction, column 4 the age in Gyr at $\rm 10^3 L_{\odot}$,
column 5 the $\rm \alpha_{mlt}$. Columns 6 and 7 are the $\rm \chi^2$ goodness of fit 
between models and data respectively in the HR diagram and the radius luminosity diagram.
$\rm \chi_1^2$ in column 6 is based on effective 
temperatures of stars of group I. It is defined by
$\rm \chi_1^2=\sum_{i=1}^{N}\frac{1}{N}[\frac{T_{eff}^{mod}-T_{eff}^{obs}}{\Delta T_{eff}^{obs}}]^2$
where $\rm T_{eff}^{obs}$ and $\rm T_{eff}^{mod}$ are respectively the 
observed effective temperature of an object and that of the model 
corresponding to the same luminosity as the 
object. Cayrel de Strobel et al.~(\cite{cayrel01}) catalog 
mostly considers the $\rm T_{eff}$ from McWilliam~(\cite{mcwilliam90}).
Accordingly our $\rm \Delta T_{eff}^{obs}$ are adapted 
from this author to $\rm \Delta T_{eff}^{obs}=$130 K.
$\rm \chi_2^2$ in column 7 is computed from radii
of the nine stars of group II according to 
$\rm \chi_2^2=\sum_{i=1}^{N}\frac{1}{N}[\frac{R_{mod}^{2}-R_{obs}^{2}}{\Delta R_{obs}^{2}}]^2$.
We do not consider the traditional $\chi^2$ formula but the reduced $\chi^2$ 
to allow a comparison of the
fit when the number of stars is changed. 
We can see that changing the sample of stars from group I to group 
II hardly changes the goodness of the fit. Columns 6 and 7 are based on the 
the stars within 0.06 dex of the metallicity given in column 2
for consistency between the composition of the models and the observations
(the nearly solar metallicity objects over groups I and II are $\sigma$ Gem, 
$\delta$ And, 31 Leo, 39 Ari). However the $\rm \chi^2$ in parentheses in 
column 6 and 7 relate the models to all the members of group I and II respectively. 
Finally column 8 is the model name and bold fonts indicate modelling inputs changes 
with respect to the (reference) model mlt1.

\begin{table*}[Ht]
  \caption[]{Characteristics of the RGB models using the MLT.}
   \centering
     \label{Tabmod1}
 $$ 
     \begin{array}{lccccccc}
        \hline
        \noalign{\smallskip}
        $$\rm M/M_{\odot} $$   &  \rm [Fe/H]   & \rm Y        &  \rm Age\,at\,10^3 L_{\odot} &  \rm \alpha_{mlt}&   \chi_1^2       &   \chi_2^2 &  {\rm Model \, name} \\
        \noalign{\smallskip}								               
        \hline										               
        \noalign{\smallskip}								               
        0.95                   &  -0.17        &      0.2582  &  11.5   &  1.98       &  3.8 \, (3.2)     & 4.0  \, (3.8)     & \rm mlt1     \\
        {\bf 0.90}             &  -0.17        &      0.2582  &  13.9   &  1.98       &  3.0 \, (2.5)     & 3.2  \, (3.1)     & \rm mlt2     \\
        0.95                   &  -0.17        & {\bf 0.2482} &  12.3   &  1.98       &  3.4 \, (2.8)     & 3.6  \, (3.4)     & \rm mlt3     \\
        {\bf 0.95}             &  {\bf 0.00}   &      0.2582  &  13.7   &  1.98       &  0.72 \, (0.95)   & 1.5  \, (1.4)     & \rm mlt4     \\
        0.95                   &  -0.17        &      0.2582  &  11.5   &  {\bf 1.68} &  0.16 \, (0.11)   & 1.1  \, (0.91)    & \rm mlt5     \\
        1.13                   &  {\bf 0.00}   &      0.2582  &   7.5   &  {\bf 1.68} &  0.078 \, (0.070) & 0.54 \, (1.3)     & \rm mlt6     \\
        \noalign{\smallskip}
        \hline
     \end{array}
 $$ 
\end{table*}

%We confuse evolutionary tracks and isochrones. This is a valid hypothesis
%because the RGB evolution is so swift that the 
%evolutionary paths (for instance in the HRD) become undistinguishable 
%from the isochrones (Salaris et al.~\cite{salaris02}).

The following remarks can be done:

{\it i)} The model mlt1 uses the solar calibrated
$\rm \alpha_{mlt}$ and reaches $\rm 10^3 L_{\odot}$ at
$\rm \approx$11.5 Gyr i.e. nearly at the
age estimated for the local Galactic disk. Its RGB track
does not fit the six coolest objects 
within their observational error bars.
The models mlt2 and mlt3 are respectively 
less massive and helium poorer than the model mlt1.
Model mlt2 is extreme in the sense that it reaches $\rm 10^3 L_{\odot}$ at $\approx$13.9
Gyr (or $\rm 10^2 L_{\odot}$ at $\approx$13.88 Gyr) which is older 
than the current age estimate of the Universe of $13.7\pm0.12$ Gyr 
(Komatsu et al.~\cite{komatsu09}).
Model mlt3 is also extreme in the sense that its helium fraction
is the Big Bang Nucleosynthesis one (Coc et al.~\cite{coc04}) and 
evidently cannot be lowered any further. 
Both models mlt2 and mlt3 are in slightly 
better agreement to the data than model mlt1 and are 
almost undiscernable from it in Fig.~\ref{fig1a}.
They demonstrate that acceptable changes in mass or 
helium fraction can hardly improve the poor fit of model mlt1.

{\it ii)} Even though group I exhibits $\rm [Fe/H]\approx -0.1$ 
on average, one can distinguish two metallicity subgroups as 
$\sigma$ Gem, $\delta$ And and 31 Leo are nearly of solar metallicity
while $\rho$ Boo, 91 Psc and $\upsilon$ Boo are within 0.06 dex
of $\rm [Fe/H]= -0.17$ (see Table\ref{observations_table}).
It is necessary to investigate the metallicity effects.
The solar metallicity $\rm 0.95 M_{\odot}$ model 
mlt4 provides a better fit to the observations than the previous ones: 
its $\rm \chi_1^2$ fit to the whole group I is 
0.95 and it is 0.72 to the subsample
of the three solar metallicity stars. Together with models mlt2 \& mlt3,
this model should also be considered 
as extreme because of its age at $\rm 10^3 L_{\odot}$: 13.7 Gyr. 
The increase in metal fraction lowers the luminosity and thus 
slows down the evolution with respect to the 
same mass model mlt1.

{\it iii)} Models with lower $\rm \alpha_{mlt}$ than the
solar value provide much better fits to the data.
The model mlt5 ($\rm \alpha_{mlt}=1.68$, 0.95$\rm M_{\odot}$ and 
$\rm [Fe/H]= -0.17$) exhibits $\chi_1^2$=0.16. It
reaches $\rm 10^3 L_{\odot}$ at 11.5 Gyr.
The model mlt6 ($\rm \alpha_{mlt}=1.68$, 1.13$\rm M_{\odot}$ and 
$\rm [Fe/H]=0$) exhibits $\chi_1^2$=0.078. It
reaches $\rm 10^3 L_{\odot}$ at 7.5 Gyr.
The pattern of the goodness of the fit is similar if one 
considers the luminosity vs. square radius instead of the HR diagram.
The $\chi_2^2$ and $\chi_1^2$ have the same orders of magnitudes 
and variations for models mlt1 to mlt4. 
The two best fits in $\chi_2^2$ are also obtained for models mlt6 and mlt5 
with 0.54 and 1.1 respectively.
In these cases however the $\chi_2^2$s remain larger than the $\chi_1^2$s.
This is mostly due to the small errorbars in square radius
as compared to the errorbars in effective temperatures
(See Fig. \ref{fig1a} and \ref{fig1b}).

{\it iv)} Recent studies have suggested that
the models with a solar calibrated $\rm \alpha_{mlt}$ can properly describe
the RGB (Alonso et al.~\cite{alonso00}; Ferraro et al.~\cite{ferraro06}; 
VandenBerg et al.~\cite{vdb08}). {\it It is likely that our different conclusion
stems from the systematic use of non-grey atmospheres with 
a low $\rm \alpha_{mlt}=0.5$ parameter.}
Actually, VandenBerg et al.~(\cite{vdb08}) rely on the 
MARCS atmosphere models where they assume the high value $\rm \alpha_{mlt}=1.80$  
of the interior. Alonso et al.~(\cite{alonso00}) and Ferraro et al.~(\cite{ferraro06})
both rely on the Krishna Swamy~(\cite{krisnaswamy66}) $\rm T-\tau$ atmosphere 
relation but they do not mention the optical depth $\rm \tau_b$ at which 
they connect the atmosphere to the interior.
In anycase the absence of a low efficiency surface convection 
region is criticable. 
$\rm \tau_b \approx 1$ would mean that the outermost convective 
regions are handled with the interior $\rm \alpha_{mlt}$ parameter
which is not supported by the Balmer lines profiles (see \S \ref{sec2}).
On the other hand $\rm \tau_b \geq 10$ would imply 
that the Krishna-Swamy $\rm T-\tau$ relation is extended down
in the convective regime, which is uncorrect as mentioned
by Krishna Swamy~(\cite{krisnaswamy66}).

\subsection{The cool edge of the RGB and the CGM model}\label{sec53}

\begin{table*}[Ht]
 \begin{center}
  \caption[]{Characteristics of the RGB models using the CGM.}
   %\centering
     \label{Tabmod2}
 $$ 
     \begin{array}{lccccccc}
        \hline
        \noalign{\smallskip}
        $$\rm M/M_{\odot} $$   &  \rm [Fe/H]   & \rm Y  &  \rm Age\,at\,10^3 L_{\odot} &  \rm \alpha_{cgm}&   \chi_1^2       &   \chi_2^2 &  {\rm Model \,name} \\
        \noalign{\smallskip}								               
        \hline										               
        \noalign{\smallskip}								               
        0.95                   &  -0.17        & 0.2582 &  11.6          &  0.77          &   6.1  \, (5.1)      & \, 7.1  (6.2)    & \rm cgm1       \\
        {\bf 0.95}             &   {\bf 0.00}  & 0.2582 &  13.8          &  0.77          &   1.1  \, (1.6)      & \, 1.8  (2.3)    & \rm cgm2       \\
        0.95                   &  -0.17        & 0.2582 &  11.8          &  {\bf 0.62}    &   0.21 \, (0.11)     & \, 0.70 (0.57)   & \rm cgm3       \\
        1.17                   &   {\bf 0.00}  & 0.2582 &   6.9          &  {\bf 0.62}    &   0.057\, (0.077)    & \, 0.40 (0.76)   & \rm cgm4       \\
        0.95{\bf \dagger}      &  -0.17        & 0.2582 &  11.5          &  {\bf 0.80}    &   7.4  \, (6.1)      & \, 9.1  (7.8)    & \rm cgm5       \\
        \noalign{\smallskip}
        \hline
     \end{array}
 $$
 \end{center}
\begin{list}{}{}
\item[] Bold fonts indicate inputs changes with respect to the (reference) model cgm1. The model cgm5 mentioned by $\dagger$ is 
built with the semi empirical atmosphere described in section \ref{sec4}.
\end{list}
\end{table*}

\begin{figure}[Ht*]
\centering
\includegraphics[angle=90,width=8cm]{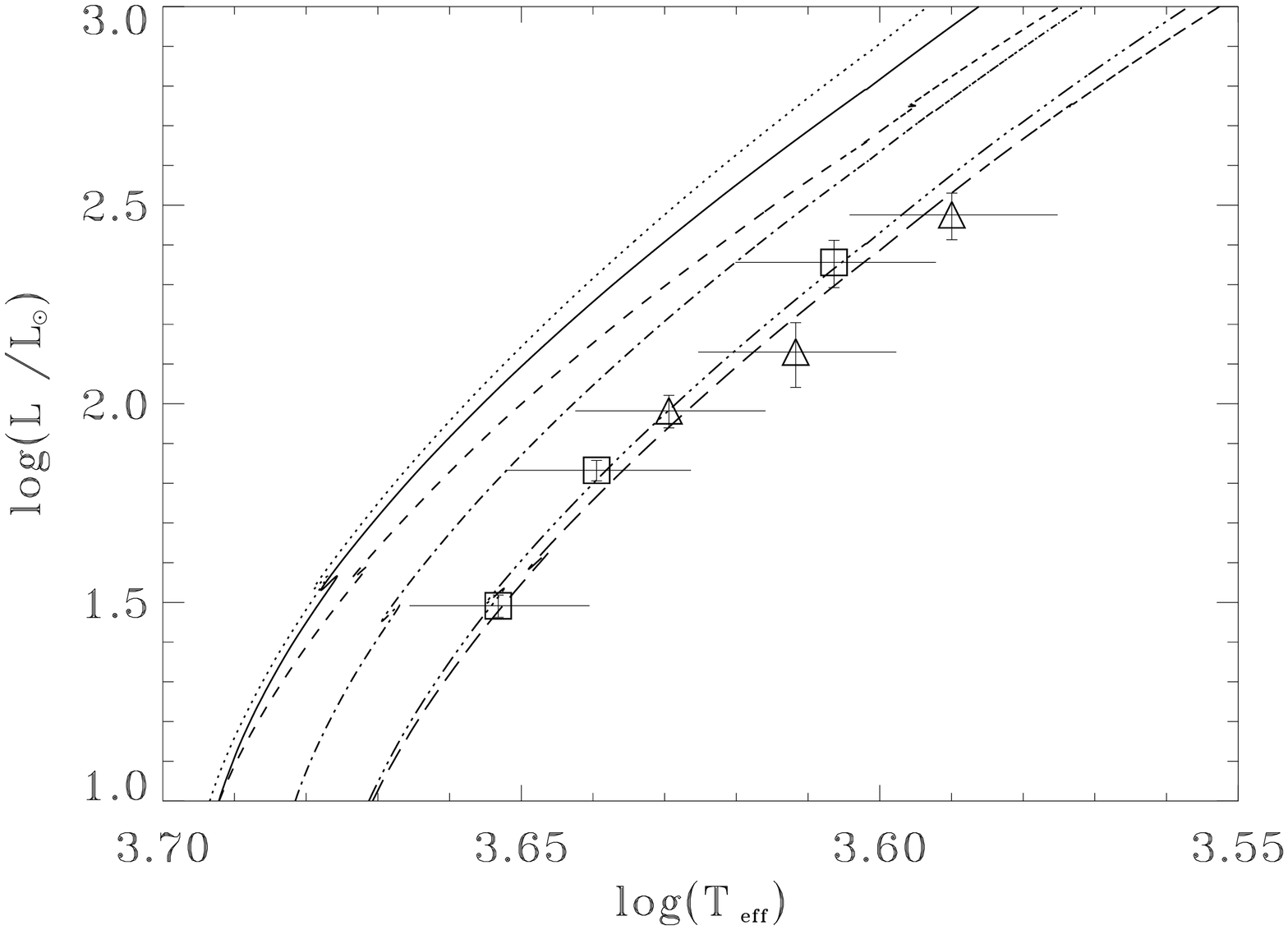}
\caption{RGBs in the HR diagram for models cgm1 to cgm5 of Table \ref{Tabmod2}.
The track of model mlt1 of Table \ref{Tabmod1} is also given 
for comparison. The solid line is model cgm1, the dashed line 
is model mlt1, the dot-dashed line is model cgm2, 
the three-dot-dashed line is model cgm3 and the long dashed 
line is model cgm4. The dotted line is model cgm5.
The data are the six giants of group I 
defined in the text and Fig.~\ref{fig0a}.
Following Fig.~\ref{fig1a} the squares and triangles 
respectively stand for solar and subsolar metallicity 
objects. Only models cgm3 and cgm4 with lower $\rm \alpha_{cgm}$ than 
the other models fit the data satisfactorily.
\label{fig2a}} % hrgb1.pro 
\centering
\includegraphics[angle=90,width=8cm]{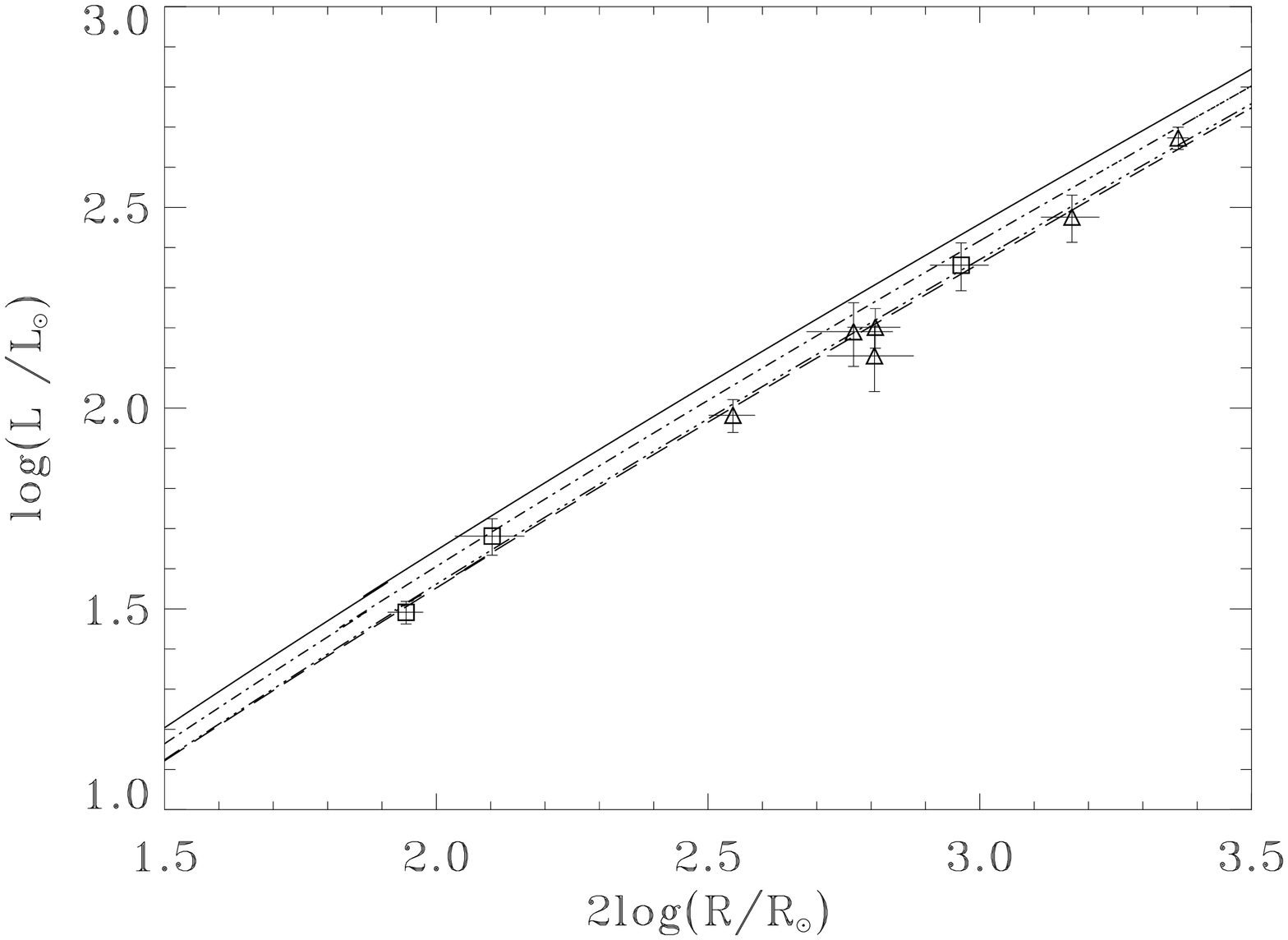}
\caption{RGBs in a bolometric luminosity vs. stellar squared radius
for models cgm1 to cgm4 of Table \ref{Tabmod2}. The lines conventions 
are similar as in Fig.~\ref{fig2a}.
The data are the nine giants of group II 
defined in the text (see also Fig.~\ref{fig0b}). 
The data symbols are the same as in Fig.~\ref{fig1b}:
Triangle symbols are for objects within 0.04 dex of solar metallicity, triangles 
for objects within 0.06 dex of $\rm [Fe/H]=-0.17$.
\label{fig2b}} % hrgb1.pro
\end{figure}

The MLT is widely used in stellar modelling.
However the FST approach initialy developped by 
Canuto \& Mazzitelli~(\cite{canuto91})
is physically more consistent (Mazzitelli~\cite{mazzitelli99}).
When compared to observations it seems to provide 
a better description of stellar convection
as supported by many recent studies in seismology 
(Basu \& Antia~\cite{basu94}; Monteiro, Christensen-Dalsgaard \& Thompson~\cite{monteiro96};
Samadi et al.~\cite{sam06}) and stellar evolution (Stothers \& Chin~\cite{stothers97}; 
Montalban et al.~\cite{montal01}).
For this reason we find it important to check possible
changes of the characteristic length scale of the FST model.
In a more appropriate version of this model, the 
characteristic length scale is the 
distance to the convectively stable region and 
therefore is not a free parameter.
This approach is not implemented in our stellar evolution code.
It requires substantial changes to compute in
the same iterations both the convection efficiency
and distance to the convectively stable region.
Instead we use the Canuto, Goldman \& Mazzitelli~(\cite{cgm96})
version (CGM) of the FST with a characteristic
length scale $\rm \alpha_{cgm}H_p$ as mentioned 
in \S \label{sec2} above. The precise description 
of the version of the model including the parameters 
is given in appendix 2. 

Table \ref{Tabmod2} sums up the properties of the
models built using the CGM model while Fig.~\ref{fig2a} and \ref{fig2b}
display their RGB evolutionary tracks. 
Columns conventions are the same in Tables \ref{Tabmod2}
and \ref{Tabmod1}.
We did not report here on the models exploring the 
effects of a change in mass or helium fraction 
(corresponding to models mlt2 \& mlt3 in \ref{sec52}). 
Similar changes of these parameters change the location 
of the RGB as little as in the case of the MLT.

The following remarks can be done:

{\it i)} Models cgm1 and mlt1 are identical except for
their atmosphere boundary conditions and convection prescriptions.
They both suffer the same drawback as the RGB they 
define are too warm with respect to the observations.
Model cgm1 RGB is even warmer than that of model mlt1. This is a consequence 
of the higher efficiency of adiabatic convection in the CGM prescription 
than in the MLT. In the case of deep convective envelopes this
produces higher effective temperature objects 
(Mazzitelli, D'Antona \& Caloi~\cite{mazzitelli95}).
As can be seen in Fig.~\ref{fig2a}, for larger luminosities (i.e. 
for larger convective envelope) the difference in effective temperatures 
between models mlt1 and cgm1 becomes larger.

{\it ii)} Having a solar metallicity, the model cgm2 is 0.17 dex
more metal 
rich than the model cgm1. However in spite of its lower effective temperature 
and larger radius it also fails to fit the corresponding 
observations. As for the mlt models, the $\chi^2$ fits
in Table \ref{Tabmod2} are related to the subgroups
of stars exhibiting the same metallicity as the models.
For the $\rm[Fe/H]=0$ models, only $\sigma$ Gem, $\delta$ And and 31 Leo
are considered within group I while only $\sigma$ Gem, 31 Leo and 39 Ari
are considered within group II.
Models with masses lower than cgm2 and the same metallicity could
in principle fit the data. However with an age 
over 13.8 Gyr at $\rm 10^3 L_{\odot}$, model cgm2 
sets a lower limit in mass.

{\it iii)} Both models cgm1 and cgm2 have the solar 
calibrated $\rm \alpha_{cgm}$ value. The sub 
solar $\rm \alpha_{cgm}$ models cgm3 and cgm4
produce a better agreement to the selected samples
of stars in the two considered diagrams (Fig.~\ref{fig2a}
and \ref{fig2b}) as can be also seen in Table \ref{Tabmod2}. 
The model cgm3 has $\rm [Fe/H]=-0.17$ and 0.95 $\rm M_{\odot}$ while 
cgm4 has $\rm [Fe/H]=0$ and 1.17 $\rm M_{\odot}$. Thus models cgm3 and cgm4 
correspond respectively to models mlt5 and mlt6 in terms of
masses and metallicities.

{\it iv)} In an attempt to investigate the atmosphere modelling effects, 
the model cgm5 was built with the semi empirical approach for the 
atmosphere calculation described in section \ref{sec4}.
Accordingly its $\rm \alpha_{cgm}$ is 0.80
which is the solar convective scale when the semi empirical 
atmosphere are used. 
The cgm5 RGB lies very close to the 
cgm1 RGB and is therefore too warm to fit the lower 
envelope of the observations. Models cgm1 and cgm5
illustrate that a change to semi empirical atmospheres
does not reduce the gap between calculations and
observations.

\subsection{The global RGB and the seismic targets}\label{sec54}

\begin{table*}[Ht]
 \begin{center}
  \caption[]{Mass repartition for the MLT and CGM models of convection.}
   %\centering
     \label{tab4}
 $$ 
     \begin{array}{lccccccc}
     \hline
     \noalign{\smallskip}
     \rm Mass\, range                            &  \rm  \alpha_{mlt} &  \rm  \alpha_{mlt}    &  \rm  \alpha_{cgm}  & \rm  \alpha_{cgm}   &   \rm Truncated   \\
                                                 &  \rm  1.98         &  \rm  1.68            &  \rm  0.77          & \rm  0.62           &   \rm      PDMF   \\
     \noalign{\smallskip}			                                           					               
     \hline					                                           					               
     \noalign{\smallskip}			                                           					               
     \rm M < 1.5 M_{\odot}                       &  76\pm4\%          &  54\pm13\%            &  84\pm15\%          &   49\pm11\%     &    54\%    \\
     \rm 1.5 M_{\odot} \leq M \leq 2.5 M_{\odot} &  24\pm8\%          &  41\pm11\%            &  16\pm7\%           &   40\pm10\%     &    27\%    \\
     \rm M > 2.5 M_{\odot}                       &  0\%               &  5\pm4\%              &  0\%                &   11\pm5\%      &    19\%    \\
     \noalign{\smallskip}
     \hline
     \end{array}
 $$
 \end{center}
\end{table*}

We now intend to use the rest of the data we have about the sample
of 38 stars. First, we check how the stars are distributed between the 
evolutionary tracks of different masses. We consider the two surface convection 
prescriptions with convective length scales calibrated on the Sun 
or on the lower envelope of the current RGB in the preceding section.

Assuming that the stellar mass distribution of our sample is 
the present day Galactic field stellar mass function (hereafter PDMF), the 
question we can investigate is whether the distribution of the sample between 
the evolution tracks of different masses agrees
with the PDMF. The use of the PDMF here relies on two assumptions. First, it assumes 
that the mass loss of the preceding main sequence and giant branch evolution is negligible. 
There is no indication of significant mass loss before the RGB tip.
When using the recent formula advocated in Catelan~(\cite{cat00}) we find that
the $\rm 1.5 M_{\odot}$ model looses less than $\rm 0.005 M_{\odot}$ before reaching
$\rm 8\,10^2 L_{\odot}$. For larger masses, the mass loss is lower. Second, it assumes that the stars of our 
sample are a well mixed Galactic field population that formed in a variety of star forming 
regions and environments. Should a substantial fraction of the stars in our sample be formed in a 
single protocluster star forming region, deviations, both in the slope and in the characteristic 
mass, between the PDMF and that of our sample could be expected (Dib et al.~\cite{dib10}).
The observed PDMF of stars in the local Galactic field is well fitted by a multi-exponent power 
law functional form (Kroupa~\cite{kroupa02} and~\cite{kroupa07}):

\begin{equation}
\rm \frac{dN}{dM}=k (0.5/0.08)^{-1.3} (M/0.5)^{-2.3}\,if \, 0.5 < M/M_{\odot} < 1.0
\label{eq2}
\end{equation}
\begin{equation}
\rm \frac{dN}{dM}=k (0.5/0.08)^{-1.3}(1.0/0.5)^{-2.3} (M)^{-2.7}\, if \, M/M_{\odot} > 1.
\label{eq3}
\end{equation}

In Eq.~\ref{eq2} and \ref{eq3}, k=0.877 and N is the density of 
stars of mass M (in solar mass units). The corresponding repartition 
below between and above 1.5 and 2.5 $\rm M_{\odot}$ is given in 
the last column of Table \ref{tab4}.
We name it 'truncated' PDMF as we derive the distribution using the previous power laws and 
making the hypothesis that no star has a mass lower than $\rm 0.95 M_{\odot}$
which we found to be the mass of the objects making the lower envelope of the RGB (\S \ref{sec52}).
If we had considered a lower truncature mass such as $\rm 0.90 M_{\odot}$ (corresponding to an unrealistic
age for the oldest stars of the sample) the distribution
would hardly have been affected.
As is well known, the position of the RGB strongly depends
on the metallicity. For instance, for the
$\rm 1.5 M_{\odot}$ CGM models at $\rm 10^2 L_{\odot}$, a change
from $\rm [Fe/H]=-0.34$ to $\rm [Fe/H]=0$ lowers the $\rm T_{eff}$ by 240 K 
(and respectively increases the radius from 15.8 to 17.7 $\rm R_{\odot}$). 
Thus the distribution of stars in regards to evolutionary tracks
is made with models nearly having the same metallicity as the observed stars.
We compare data and models for three metallicities. The $\rm [Fe/H]=-0.34$
tracks are used for objects exhibiting $\rm [Fe/H]<-0.25$.
The $\rm [Fe/H]=-0.17$ tracks are used to estimate the distribution of objects 
having $\rm -0.25 \leq [Fe/H] \leq -0.08$. Finaly the $\rm [Fe/H]=0$ tracks are 
used when the observation gives $\rm [Fe/H]>-0.08$.

\begin{figure}[Ht]
\centering
\includegraphics[angle=90,width=8cm]{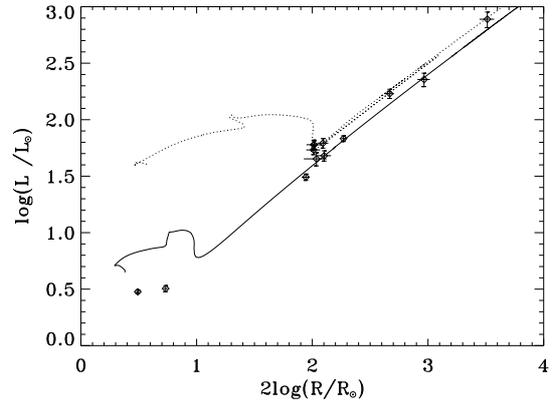}
\caption{Exemple of RGBs in the bolometric luminosity vs. stellar squared radius
diagram for models using the CGM
convection model and having $\rm [Fe/H]=0$: 1.5 $\rm M_{\odot}$ solid line 
and 2.5 $\rm M_{\odot}$ dashed line. The evolutionary 
paths are shown from the zero age main sequence. For the sake of clarity
only the stars with $\rm [Fe/H]>-0.08$ are displayed.\label{fig3}} % hrgb1.pro 
\end{figure}

The repartition between the tracks has been drawn from the luminosity vs. 
stellar squared radius diagram (see Fig. \ref{fig3}).
It is reported in Table \ref{tab4} for the different assumptions on the 
convective length scales. Column 2 shows the distribution for 
stars with $\rm \alpha_{mlt}=1.98$, column 3 for stars with $\rm \alpha_{mlt}=1.68$,
column 4 for stars with $\rm \alpha_{cgm}=0.77$ and column 5 for stars with 
$\rm \alpha_{cgm}=0.62$. For a given mass, composition and
luminosity, the larger the characteristic convective length scale the warmer
the model. Being warmer, the solar calibrated models ($\rm \alpha_{mlt}=1.98$ or $\rm \alpha_{cgm}=0.77$)
suggest a local RGB strongly biased toward lower mass stars. The models
calibrated on the lower envelope of the RGB 
($\rm \alpha_{mlt}=1.68$ or $\rm \alpha_{cgm}=0.62$) are in 
better agreement with the PDMF. This confirms the slightly 
less efficient CGM convection in the RGB regime found in
section \ref{sec53}. However this trend is merely indicative.
The small sample of stars in each mass bins induces large statistical 
uncertainties. Moreover the RGB tracks of stars with 
convective cores on the main sequence depend on 
the amount of overshooting during this phase. This amount
of core overshooting is not a well established quantity.
The larger the overshoot the more massive the helium cores and the
smaller the radius at a given luminosity. According
to the mass dependence of core overshooting of Claret~\cite{claret07} (see Fig. 13 of this author),
we have taken $\rm \alpha_{ov}=0.1 H_p$ and $\rm \alpha_{ov}=0.2 H_p$
in $\rm 1.5 M_{\odot}$ and $\rm 2.5 M_{\odot}$ models respectively.

\begin{figure}[Ht]
\centering
\includegraphics[angle=90,width=8cm]{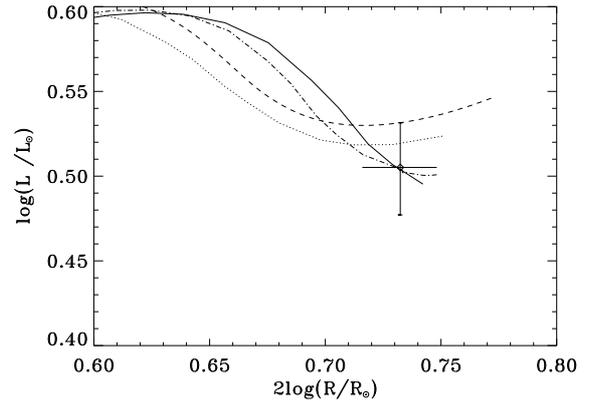}
\caption{$\delta$\, Eri evolution tracks for $\rm \alpha_{cgm}=0.62$
(solid line) $\rm \alpha_{cgm}=0.77$ (dotted line), $\rm \alpha_{mlt}=1.68$ 
(dot-dashed line) and  $\rm \alpha_{mlt}=1.98$ (dotted line). All models
match the observed radius at the observed luminosity within the errorbars.
However the lower convective length scales models $\rm \alpha_{cgm}=0.62$ 
and $\rm \alpha_{mlt}=1.68$ better fit the data. \label{fig4}} % hrgb1.pro 
\end{figure}

\begin{figure}[Ht]
\centering
\includegraphics[angle=90,width=8cm]{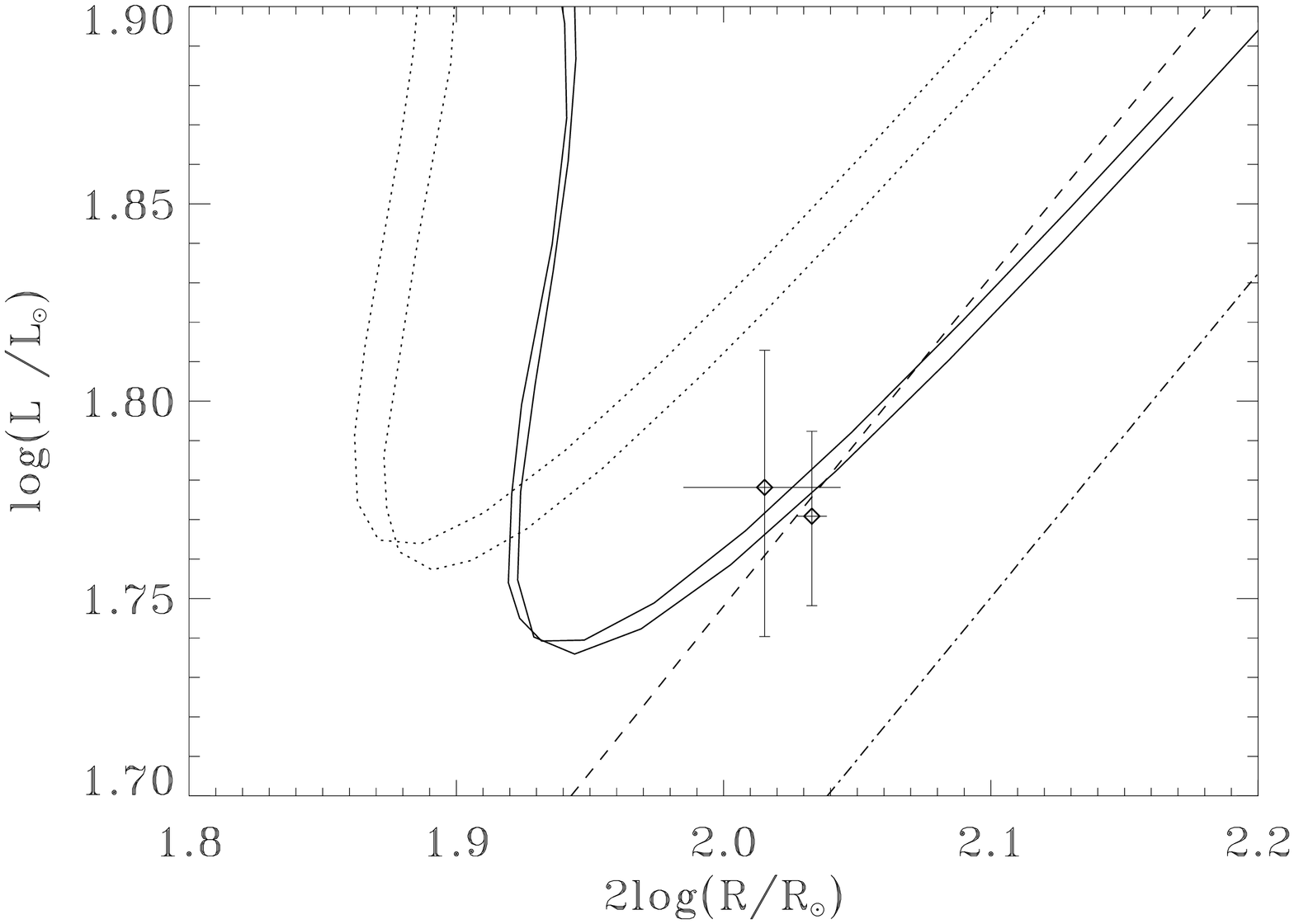}
\caption{Solid lines show the evolution tracks for $\rm \alpha_{cgm}=0.62$
and $\rm \alpha_{mlt}=1.68$ and a mass of $\rm 2.65 M_{\odot}$.
The dotted lines correspond to the same mass but $\rm \alpha_{cgm}=0.77$
and $\rm \alpha_{mlt}=1.98$. The solar calibrated models
for convection do not fit the observations. Dashed line 
and dot-dashed lines are $\rm 1.85 M_{\odot}$ models respectively
for $\rm \alpha_{cgm}=0.77$ and $\rm \alpha_{cgm}=0.62$. 
The lower radius object is $\xi$\,Hya and the larger $\epsilon$\,Oph. 
\label{fig5}} % hrgb1.pro 
\end{figure}

We dedicated a special attention to $\delta$\,Eri, $\xi$\,Hya and $\epsilon$\,Oph
as asteroseismology has allowed an accurate determination of their masses. 
Our $\delta$\,Eri models have a mass of 1.215 $\rm M_{\odot}$ (Th\'evenin et al.~\cite{thevenin05}) and
X=0.7256 to account for the star's metallicity $\rm [Fe/H]=0.13$.
The $\epsilon$\,Oph models have a mass of 1.85 $\rm M_{\odot}$ (Mazumdar et al.~\cite{mazumdar09}) and
X=0.7326 ($\rm [Fe/H]=-0.12$). Finaly the $\xi$\,Hya models
have a mass of 2.65 $\rm M_{\odot}$ (Th\'evenin et al.~\cite{thevenin05}) and a hydrogen mass fraction 
X=0.7308 ($\rm [Fe/H]=-0.04$). All those models
have the same helium mass fraction Y=0.2582. 
We assume a convective core overshooting of 0.1 $\rm H_p$
in $\delta$\,Eri and 0.2 $\rm H_p$ in $\xi$\,Hya and $\epsilon$\,Oph.
Fig.\ref{fig4} shows that the position of the $\delta$\,Eri models weakly
depends on the chosen convection length scale.
The $\rm \alpha_{cgm}=0.77$ and $\rm \alpha_{mlt}=1.98$ 
(solar calibrated) models cannot be ruled out as they 
fit the data in the errorbars. However
the $\rm \alpha_{cgm}=0.62$ and $\rm \alpha_{mlt}=1.68$ 
models converge right to the observed radius at the observed luminosity!
For $\xi$\,Hya only the $\rm \alpha_{cgm}=0.62$ and 
$\rm \alpha_{mlt}=1.68$ models fit the
observations within the errorbars: Fig.\ref{fig5}. On the contrary, the
$\epsilon$\,Oph tracks with the low $\rm \alpha_{cgm}=0.62$ does not 
fit them. It is a very interesting point. The explanation
might be that $\epsilon$\,Oph is not on its first ascent of the RGB
but a 'red clump' star i. e. in the helium core burning phase. 
There are several clues of this. First $\epsilon$\,Oph
belongs to the tight group of six stars that
can be seen at $\rm log(L/L_{\odot})\approx1.7$ on 
Fig.\ref{fig0b} and that strongly suggest the red clump.
The second clue comes from Mazumdar et al.~(\cite{mazumdar09}) models. 
They spend 20 times more time in $\epsilon$\,Oph errobox in the 
HR diagram while on the helium core burning stage
than while on the the first ascent of the RGB
which means it is much more likely that we observe an helium 
burning $\epsilon$\,Oph \footnote{Our stellar evolution 
code cannot pass the so-called helium flash and follow 
the subsequent evolution of stars lighter than $\rm \approx 2.1 M_{\odot}$
like $\epsilon$\,Oph during their helium core burning phase.}.
Finaly, except for its mass, $\epsilon$\,Oph seems identical to $\xi$\,Hya
which is likely on its first ascent of the RGB.
Their metallicities only differ by 0.08 dex. 
Their luminosities ($\rm 5 \%$ difference), radii ($\rm 2 \%$ difference) and 
effective temperatures ($\rm 80 K$ difference)
are the same within errorbars but the masses strongly differ.
Such a difference induces a difference in models radii of $\rm \approx 15\%$
at $\rm \approx 60 L_{\odot}$ if both stars are on the RGB and if 
the same convective length scale is taken. 
Contrary to $\epsilon$\,Oph, $\xi$\,Hya is probably not 
in the helium burning
stage. We found the minimum luminosity of the 2.65 $\rm M_{\odot}$ models
during the core helium burning (eAGB) around 80 $\rm L_{\odot}$. 
This value, in agreement
with stellar grid calculations (Schaller et al.\cite{schaller92}), stands 
well outside the luminosity errorbar for $\xi$\,Hya.
Finaly, leaving apart the case of $\epsilon$\,Oph, the seismic data 
combined to radii measurements suggest a drop in the 
convective length scale confirming the previous results. 
The seismic targets are in small number at the moment
and Fig.\ref{fig4} and \ref{fig5} mostly illustrate that 
combining seismologically determined masses to interferometric 
radii offers very sensitive tests of the surface convection
efficiency. In the near future seismology should allow many more
accurate mass and radius determinations (Basu, Chaplin \& Elseworth~\cite{basu10}).

% La distribution de type stellaire au voisinage solaire est sur les
% type tardifs (demander Sami : lois Chabrier \& Kroupa). 
% On attend une loi fortement biaisée vers les faibles masses. On exclu
% les étoiles moins massives que XX Mo (modèle diffusif) étant
% donnée leur temps d'évolution et l'âge de l'Univers (Reférence WMAP). 

% Les tracés évolutifs le long de la branche des géantes dépendent
% peu de la masse (Référence Salaris). A 100 Lo la différence de Teff est : XX K
% Cependant on peut mettre une limite supérieure sur la masse.
% Si la composition est déduite de la valeur moyenne [Fe/H]=-0.18 
% X=, Y=  (cf section suivante) la figure 2a suggère une gamme de masse..
% Commenter sur les étoiles sous la RGB.
% Si la composition est solaire figure 2b la gamme de masse est...

% Commenter sur l'outlier et sur les deductions spectroscopiques de M 
% grace à log g.

\section{Discussion}\label{sec6}

This work aims at constraining the outer convection 
prescriptions in low mass red giants and subgiants by using
their absolute luminosities, effective temperatures,
interferometric measurements of their radii and seismic data.
The observational sample is made out of 38 Galactic disk nearby
stars. It was selected on the basis of interferometric
radii measured with a better than 10 percent 
accuracy. The average metallicity of the sample 
is $\rm [Fe/H]=-0.17$. There are small but significant metallicity
differences between the stars.
Age and mass differences are expected as well.
From the modelling point of view, we used a modified version of the secular 
stellar evolution code CESAM. 
A special care was devoted to the atmosphere boundary 
conditions. We computed two grids of non grey atmospheres
surrounding the expected surface gravities and
effective temperatures. The first grid, based
on the PHOENIX/1D calculations, relies on the mixing
length theory for the treatment of convection while the
second grid, based on the ATLAS12 calculations, relies
on a modified version of the full spectrum of turbulence model
(see appendix 2).
We considered the boundary conditions to the interior model
at the Rosseland optical depth 20.
In the regime of surface conditions encountered in red giants,
superadiabatic convection extends above and below
this limit. The same model of convection was used
for the atmosphere and the interior but 
different convection length scales were adopted in the two regions.
A procedure of linear interpolation
of the thermal gradient with the optical depth allows a smooth 
transition of it between them. Other important changes to CESAM 
were made to enable modeling of late stages of stellar 
evolution (see appendix 1). We chose the Asplund et 
al.~(\cite{aspl05}) solar repartition for metals 
in the opacity tables. The total
metallicity was varied to account for solar or slightly
sub solar metal content in the sample and the  
variations between the stars in dedicated cases.

We proceeded in three steps. First we built 
calibrated solar models for the two prescriptions 
of convection: the MLT and the CGM version of the full spectrum of 
turbulence. Then we built RGB models for different masses, metallicities
and initial helium fractions. The comparison
between models and data was made in the classical HR diagram
and in the squared radius vs. luminosity
diagram. In the third step we investigated the mass
distribution on the red giant branch.
To that extent we made the assumption that the
mass distribution of the data set follows the 
present day Galactic field stellar mass function.
Finaly we use of the asteroseismically
determined masses in the few cases where they 
are available.

Our conclusions are the following:

i) Lower envelope of the RGB

When the mixing length theory is used
with the solar convective length scale ($\rm \Lambda=\alpha_{mlt}H_p$ 
and $\rm \alpha_{mlt}=1.98$), 
the model with $\rm 0.95 M_{\odot}$ \& $\rm[Fe/H]=-0.17$ 
ascends the RGB at $\approx$ 11.5 Gyr i.e. the
estimated age of the local Galaxy disk.
This model poorly fits the lower envelope of the RGB in effective temperatures
or radii. The situation is not improved by lowering of 
the mass or the helium fraction unless we consider models older
than the Universe or having an helium fraction below the Big
Bang Nucleosynthesis value.
Considering a solar metallicity model of $\rm 0.95 M_{\odot}$ 
improves the fit but at the expense of an age (13.7 Gyr) 
hardly compatible with current constraints on the age of the Galaxy.
Models with lower convective length scale ($\rm \alpha_{mlt}=1.68$)
much better fit the observations both in 
effective temperatures and radii.  
Those models suggest two age and mass components
for the lower envelope of the local RGB. The $\rm 0.95 M_{\odot}$ 
\& $\rm[Fe/H]=-0.17$ model reaches $\rm 10^3 L_{\odot}$
at 11.5 Gyr and the $\rm 1.13 M_{\odot}$ \& $\rm[Fe/H]=0$ model reaches 
this luminosity at 7.5 Gyr. Interestingly this result
is in agreement with the work of Liu \& Chaboyer~(\cite{liu00}) who 
estimate that the age of field stars showing $\rm[Fe/H]=-0.18$ is
$\rm 11.7\pm1.9$ Gyr and the age of field stars with $\rm[Fe/H]=0$
is $\rm 7.5\pm1.7$ Gyr (see the Table 5 of these authors).

When the full spectrum of turbulence model is used
with the solar convective length scale ($\rm \Lambda=\alpha_{cgm}H_p$ and 
$\rm \alpha_{cgm}=0.77$), the $\rm 0.95 M_{\odot}$ 
\& $\rm[Fe/H]=-0.17$ model is also too warm 
with respect to the lower envelope of the red giant branch.
Correspondingly it gives too small radii when compared to
the observed ones. This feature is expected as
when the convection is efficient (e.g. in giant stars envelopes), 
the CGM model convective flux is roughly ten times larger than with the 
MLT one (Canuto, Goldman \& Mazzitelli~\cite{cgm96}). Eventually this produces higher 
effective temperatures. One needs a convectively less efficient model $\rm \alpha_{cgm}=0.62$,
with $\rm 0.95 M_{\odot}$ \& $\rm[Fe/H]=-0.17$ to provide a good fit whether 
the surface temperature or the radius is used.
Like in the MLT case the CGM models suggest two
ages for the cool edge of the local RGB, the low
metallicity component ($\rm[Fe/H]=-0.17$) being 11.8 Gyr 
and the high metallicity one ($\rm[Fe/H]=0$) being 6.9 Gyr.
We did not test the more appropriate version 
of the full spectrum of turbulence
with the convective length scale $\rm \Lambda$ equal to the distance z 
to the convective boundary.

ii) Global RGB and seismic targets

We studied the mass repartition on the local RGB
using 1.5 and 2.5 $\rm M_{\odot}$ models.
We found that models having the solar calibrated convective
lengths predict a mass distribution on the
RGB that seems incompatible with the local stellar mass distribution function.
The lower $\rm \alpha_{mlt}=1.68$ and $\rm \alpha_{cgm}=0.62$ 
are in better but not complete agreement 
with the expected mass distribution.
Stars with seismic mass determination offer a final way of probing
surface convection efficiency. Only two stars of 
the sample could be used that way. The models dedicated
to them are in excellent agreement with the constraints 
on luminosities and radii. Confirming the previous results,
they support a slightly lower than solar surface convection 
efficiency on the RGB.

To describe the RGB the convective length scale needs
to be slightly decreased with respect to the Sun in the versions of the
MLT and the CGM we used. This result is suggested 
independently by effective temperatures, radii measurements
and the few constraints from asteroseismology.
The drop is in disagreement with previous analyses
on the red giant branch. We interpret this point as a 
consequence of our systematic use of atmosphere models
with low efficiency convection. In surface layers, the specific 
entropy increases inwards from the point where 
the convection sets on until the region where the convection 
becomes adiabatic. In a solar model 35\% of this increase occurs 
above the optical depth $\rm \tau=20$. 
But on the RGB typically 15\% of the increase occurs 
above $\rm \tau=20$ (estimate from model cgm4 at 100 $\rm L_{\odot}$). 
Going from solar to RGB surface conditions means that
the superadiabatic convection -and therefore the radius of the star-
depends less and less on the atmosphere.

%Les mesures en bande K n'affectent pas le rayon mesure si teff> 4000K.
%On compare les incertitudes sur alpha liees a la methode photometrique 
%et a la methode angulaire.

% La MLT avec la calibration solaire ne s'applique pas à la RGB. Cela est en accord avec
% le travail de Mazzitelli: la FST donne de meilleurs résultats (cas de M92, analyse Mazzitelli et al. 1995).

% Questions & answers
%
% The solar calibrated models MLT and FST reproduce the lower edge of the RGB?
% Note : different solar calibrated models produce different preMS and postMS
% evolutions in Montalban et al. 2004
%
% Do we need to change the MLT alpha and FST alpha from the Sun to the RGB? 
% How is the convection efficiency evolving? Montalban suggests it decreases
% from the Sun to preMS.
% How dependent on the models does the RGB location get? 
%
% The version of the CGM we use is less more efficient than the Lambda=z version.

\begin{acknowledgements}
L. Piau is indebted to the anonymous referee 
whose remarks really helped improving the quality
the work. He also thanks F. Kupka and R. Samadi for their
help in implementing the CGM grid of atmosphere models to CESAM.  
L. Piau is member of the UMR7158 of the CNRS.
This work was supported by the French \emph{Centre National de la Recherche Scientifique, CNRS\/} 
and the \emph{Centre National d'Etudes Spatiales, CNES\/}.
S. Dib aknowledges partial support from the MAGNET project of the ANR.
This work received the support of PHASE, the high angular resolution
partnership between ONERA, Observatoire de Paris, CNRS and University Denis Diderot Paris 7.
This research took advantage of the SIMBAD and VIZIER databases at the CDS, Strasbourg (France),
and NASA's Astrophysics Data System Bibliographic Services.
\end{acknowledgements}
\appendix

\section*{Appendix 1: new CESAM integration variables}\label{ap1}

It is worth mentioning an important change 
in the integration scheme of the stellar evolution code
we use.
The hydrostatic equilibrium equation ($\frac{dp}{dm}=-\frac{Gm\rho}{r^2}$)
and the radius-mass equation ($\frac{dr}{dm}=\frac{1}{4\pi \rho r^2}$)
have singularities at the
center of a star. For this reason and also because
the CESAM code uses piecewise polynomials to describe the stellar 
structure (Morel~\cite{morel97}), the
integration variables chosen initialy 
are $\rm m^{2/3}$, $\rm \xi=r^2$ and $\rm \lambda=\ell^{2/3}$ where m, r and $\ell$
are respectively the local lagrangian
mass coordinate, radius and luminosity.
This choice avoids the singularities and makes the calculations stable.
However it prevents the calculation of the stellar structure
in case the luminosity becomes negative. 

Such a situation arises in the helium cores of low mass stars when they 
ascend the red giant branch.
Then the core is devoid of any energy source (but its own slow contraction)
but efficiently looses energy because of the neutrinos.
At some point the hottest region in the star becomes
a spherical shell which generates an inward radiative 
energy flux below i.e. a  negative luminosity.
We changed the integration variables of the CESAM code
to $\rm m^{1/4}$, $\rm \xi=r^2$ and $\rm \lambda=\ell$. This 
choice leaves no singularities is stable and allows
the calculation of the structure even though the 
luminosity becomes negative. To our knowledge this choice of integration 
variables has not been made with
the CESAM code in previous calculations.

\section*{Appendix 2: the current FST convection treatment}\label{ap2}

There are several versions of the MLT as well as of the FST model.
Whenever such treatments of convection are used 
it is important to describe them and to provide their parameters
in order to make precise comparisons to other works possible.
The current MLT convection treatment is described in detail
in Piau et al.~(\cite{piau05}) whereas the current FST version 
relies on the Canuto, Goldman \& Mazzitelli~(\cite{cgm96}) 
equations. The same equations are used both for the 
atmosphere and the interior convection.
The atmosphere models were computed with the ATLAS12 code (Castelli~\cite{castelli05})
and especially for the actual solar surface 
composition i.e. X=0.7392, Z=0.0122 and 
the metal repartition advocated by Asplund et al.~(\cite{aspl05}).
The convective flux $\rm F_{conv}$ is given according to:
$$\rm F_{conv}=K_{rad}TH_p^{-1} (\nabla-\nabla_{ad})\Phi (S)$$
where $\rm K_{rad}=\frac{4acT^3}{3\kappa \rho}$ is the radiative
conductivity and S is the convective efficiency:
$\rm S=Ra \times Pr \propto \Lambda^4$. Ra and Pr are
respectively the Rayleigh and Prandtl numbers 
of the convective flow while $\rm \Lambda$ is its 
characteristic length scale. All other symbols
keep their traditional meanings.

The $\rm \Phi (S)$ function is the ratio of convective
to radiative conductivity and is given by:
$$\rm \Phi (S)=F_1(S)F_2(S)$$
where 
$$\rm F_1(S)=(K_o/1.5)^3 a S^k((1+bS)^m-1)^n$$ 
and 
$$\rm F_2(S)=1 + \frac{cS^p}{1+dS^q} + \frac{eS^r}{1+fS^t} $$
with
$\rm K_o$=1.7, the Kolmogorov constant, a=10.8654,
b=0.00489073, k=0.149888, m=0.189238, n=1.85011,
c=0.0108071, p=0.72, d=0.00301208, q=0.92, e=0.000334441, 
r=1.2, f=0.000125, t=1.5.

The equations used here are similar to those of
Heiter et al.~(\cite{hei02}). However unlike those
authors we do not adopt $\rm \Lambda=z$ where 
z is the distance to the boundary between radiatively
stable and unstable regions. It would require substantial
changes of our code to solve the stellar structure
equations with $\rm \Lambda=z$. The reason is that this
distance z is not known before the equations have been 
solved and makes the problem non local. 
Instead $\rm \Lambda=\alpha_{cgm}H_p$ is a local quantity
which leads
to a simpler integration scheme.

\end{document}